\documentclass[a4paper,fleqn,usenatbib]{mnras}
\usepackage[usenames, dvipsnames]{color}
  
\usepackage{graphicx}	% Including figure files
\usepackage{amsmath}	% Advanced maths commands
\usepackage{amssymb}	% Extra maths symbols
\title[Mass from a third star]{
Mass from a third star: transformations of close compact-object binaries within hierarchical triples} 
\author[R. Di~Stefano]{R. Di~Stefano$^{1}$
\thanks{E-mail: rdistefano@rcfa.harvard.edu}
\\
$^{1}$Harvard-Smithsonian Center for Astrophysics, 60 Garden St, Cambridge, MA 02138,  US\\
} 

\date{Accepted XXX. Received YYY; in original form ZZZ}

\pubyear{2018}

\begin{document}
\label{firstpage}
\pagerange{\pageref{firstpage}--\pageref{lastpage}}
\maketitle

\begin{abstract}  
Close-orbit binaries consisting of two compact objects are a 
center of attention because of the detection of 
gravitational-radiation-induced mergers. The creation of
close, compact-object binaries involves 
physical processes that are not yet well 
understood; there are open questions about the manner in which 
two compact objects come to be close enough to merge within a Hubble time.
Here we explore an important, and likely common physical process:
mass transfer from a third star
in a wider, hierarchical orbit.    
Mass added to the close binary's components can reduce the time to merger and can
even change the nature of an accretor, transforming a white dwarf to a neutron star
and/or a neutron star to a black hole. Some accreting WDs in close binaries may even explode as
Type Ia supernovae. Given the ubiquity of higher-order multiples, the evolutionary
channels we lay out may be important pathways to gravitational mergers, including Type Ia
supernovae. Fortunately, these pathways also lead to testable predictions.    

\end{abstract}
\begin{keywords}
keyword1 -- keyword2 -- keyword3
\end{keywords}

%%%%%%%%%%%%%%%%%%%%%%%%%%%%%%%%%%%%%%%%%%%%%%%%%%

%%%%%%%%%%%%%%%%% BODY OF PAPER %%%%%%%%%%%%%%%%%%

\section{Why Mass Transfer to Compact Binaries Is Important}

The discovery of gravitational radiation from the mergers of black holes (BHs)
and neutron stars (NSs) 
has made it important to understand how two stellar remnants
can come to be in a close-enough 
orbit that they will merge within a Hubble time 
\citep{2016PhRvL.116f1102A,
2017PhRvL.119p1101A,
2017PhRvL.119n1101A,
2016PhRvX...6d1015A}. 
Here we consider the
effects of mass transfer from a third star in a wider orbit. Mass gained
by components of the inner binary can change white
dwarfs (WDs) into NSs, or NSs into BHs.
Whether or not the natures of its components are altered,
modifications of the total mass and angular 
momentum of the inner binary changes the time to gravitational merger. 
For wide ranges of physically reasonable parameters, the times to 
merger are decreased, although times to merger can also increase. This 
has implications for the 
rates of formation and the rates of mergers of binaries producing
gravitational radiation, as well as for the rates of Type Ia supernovae
(SNe~Ia) and the accretion-induced collapse (AIC) of WDs to NSs and NSs to BHs.
Mass transfer from a companion in a wider orbit also
produces directly detectable signatures. These systems can be be bright, with
X-ray luminosities on the order of the Eddington luminosity. Modulations of the
X-ray luminosity on times governed by the orbital period of the 
compact binary, and possibly even binary self-lensing, can lead to
definitive identifications.

The components of the inner binary have already evolved, yet they are in 
a close orbit.
They therefore have likely 
experienced at least two epochs of prior interaction involving the transfer of mass and/or
episodes during which the binary was engulfed by a common envelope.  
The wide-orbit companion star has not yet become
a stellar remnant. 
As it evolves, it will begin to transfer mass to the inner binary during
an epoch that may be as short as $10^5$~years, or may be longer than 
$\sim 10^8$~years.

In \S 2 we develop the model and sketch key elements of the 
basic science.  
In \S 3 we present a set of
examples. 
We find that the times to mergers 
and the masses
of the compact objects are increased   
under a broad range of physically-motivated input assumptions. 
In \S 4 we focus on the transformations that are possible [e.g., through 
accretion-induced collapse (AIC) or Type Ia supernova (SN~Ia)] when mass is added to a WD or NS.  
A broader range of possibilities for the underlying physical processes is discussed in \S 5. 
In \S 6 we focus on the implications for gravitational mergers,  
collapsing WDs and NSs, and for exploding WDs. We find that hierarchical triples
may contribute to the rates of NS-NS, NS-BH, and BH-BH mergers,
as well as to the rate of Type Ia supernovae (SNe~Ia).

\section{The Model} 

\subsection{Overview}

Our model starts with a hierarchical triple. The triple contains 
a binary composed of two stellar remnants in a close orbit,
and a third, unevolved star in a wider orbit.
Although we are interested in gravitational mergers,
the two stellar remnants need not have an orbital separation  
small enough to allow them to merge within a Hubble time. 
This is because interaction with mass from the  
third star can decrease the time to merger. 
If one or both of the compact objects is a NS, the eventual merger
could nevertheless be a BH-BH merger, since a NS may gain enough mass to 
collapse. Similarly, WDs may be transformed into NSs or, with more
significant mass increase, even to BHs. 
We therefore consider inner
binaries with the full range of compact-object combinations.

The inner binary has evolved into two compact objects.  
We concentrate on the epoch during which the third star is able to 
lose mass which comes under the gravitational influence of the inner 
binary.\footnote{There may have been an earlier epoch during which 
three-body dynamics played a role.} 
 Note that, if star~3 is massive, it 
could start transferring mass during its giant phase very soon after the formation of the 
inner binary. If, on the other hand, it is a solar-mass star, the 
wait time could be billions of years.

Mass transfer takes place over a time $t_{mt}$. The value 
of $t_{mt}$ also depends on the mass of star~3. It can 
range from $10^5$~years for massive donors to tens of millions of 
years for donors of lower mass.
During the epoch of mass transfer, the characteristics of the inner 
binary can change. Specifically, the masses and orbital separation 
of the components can be altered, thereby changing the time to merger. 
As matter accretes onto the components of the inner binary, X-rays
will be emitted, When the  accretion rates is large, the system can be highly luminous.
This means that these systems can be detected as X-ray sources even in external 
galaxies. Furthermore, if the count rate is high enough, subtle effects,
such as modulation of the X-ray flux at harmonics of the inner orbit,
may be detectable.

\subsection{Orbital dimensions}

We consider a binary composed of two compact objects with masses 
$M_1$ and $M_2$, orbiting each other with semimajor axis $a_{in}$.  
If the orbit is circular, the time to merger
is
\begin{equation}  
\tau_{merge} = \Bigg(\frac{1.5\times 10^8 {\rm yr}}{M_1 M_2 (M_1+M_2)}\Bigg) \,
\Bigg[\Bigg(a_{in}^4 - a_{min}^4\Bigg)\Bigg],  
\end{equation}  
where $a_{min}$ is the separation at the time of merger, and 
masses and distances are expressed in solar  units.  
In the cases we consider, $a_{min}$ is small enough relative to
$a_{in}$ that it can be neglected in the calculations described below. 

A third body, a non-degenerate star with mass $M_3$,
is in orbit with the binary. Dynamical stability requires that 
the closest approach between $M_3$ and either component
of the binary be much larger than $a_{in}$. 
We define $q=M_2/M_1,$ with $M_2 < M_1$, and $q_{out}= 
(M_1 + M_2)/M_3$.
\citet{1995ApJ...455..640E} have derived an expression   
for the minimum possible radius, $a_{out}^{min}$ of the outer orbit. 
\begin{equation}
a_{out}^{min} = a_{in} \times \Bigg[ \frac{3.7}{q_{out}^\frac{1}{3}}  + 
\frac{2.2}{1 + q_{out}^\frac{1}{3}}  +
\Bigg(\frac{1.4}{q_{out}^\frac{1}{3}}\Bigg)\, 
\Bigg(\frac{q_{out}^\frac{1}{3}-1}{q_{out}^\frac{1}{3}+1}\Bigg) \Bigg]  
\end{equation}
In principle, $a_{out}^{min}$ could correspond to the periapse
of a wide elliptical outer orbit. Here, for the sake of simplicity and also 
because many mass transfer systems have been tidally circularized, we  
consider circular orbits. 

The finite size of the star in the outer orbit places additional
 restrictions on the 
outer orbit.
As we discuss in the appendix, the Roche-lobe picture of channeled mass
transfer can be applied when mass is transferred to a compact inner binary
from a donor in a much wider orbit.
If we assume that the triple system is in dynamical equilibrium sometime
before mass transfer from the outer star begins, then the outer star
must have fit inside its Roche lobe. Thus,  
\begin{equation} 
a_{out} > \frac{R_3}{f(1/q_{out})},  
\end{equation} 
where $f(x)=0.49\, x^{0.67}/(0.6\, x^{0.67}+lg(1+x^{0.33}))$.  
The true value of the minimum separation
between star~3 and the center of mass of the inner binary is therefore
\begin{equation} 
a_{out}^{min} = max\Bigg[a_{out}^{min} , \frac{R_3}{f(1/q_{out})}\Bigg].
\end{equation} 
 On the giant branch, the radius, $R$, and luminosity, $L$, of the star are strong functions of the instantaneous value of the core mass, $C(t)$. 
\begin{equation} 
R = 0.85\, M(0)^{0.85} + \frac{3700\, C(t)^4}{1 + C(t)^3 + 1.75\, C(t)^4}
\end{equation}
\begin{equation} 
L = M(0)^3 + \frac{10^{5.3} C(t)^6}{1 + 10^{0.4} C(t)^4 + 10^{0.5} C(t)^5}
\end{equation}
The expressions for $R$ and $L$ each consist of a first term meant to correspond to the value of the radius and luminosity, respectively, of a main sequence star. Depending on the specific value of the initial mass, expressions which differ from those above may be more appropriate. For giants, however, these terms are dwarfed by the second terms, which depends only on the core mass. Thus, the radius and luminosity of giants depends only weakly on the initial mass.

\subsection{The Flow of Mass}

Mass flows from star 3. A fraction, $\gamma,$ of $\dot M_3$ falls toward the
inner binary and the rest, $(1-\gamma)\, \dot M_3$ exits the system.
The paths taken by mass falling toward the binary may be complex. The upshot is
simply that a fraction of the incoming mass is retained by one star and 
another fraction is retained by the second star. Let $\beta_1$ and 
$\beta_2$ be the fraction of the 
mass retained by stars 1 and 2, respectively. Then
$\dot M_1 = \beta_1\, \gamma\, \dot M_3$, 
$\dot M_2 = \beta_2\, \gamma\, \dot M_3$, 
$\dot M_{in} = (\beta_1 + \beta_2)\, \gamma\, \dot M_3$. 
The remainder of the mass exits the system.

\subsection{Retention of Mass}

Many factors, including the geometry and dynamics of the mass flow,
and the action of magnetic fields determine how much infalling matter 
can be retained by a compact object. For each type of compact object we 
select a formula with which to compute $\dot M_{min}$, the minimum
accretion rate for which all mass is retained ($\beta=1$) and 
$\dot M_{max}$, the maximum accretion rate for which $\beta=1$.
For rates lower than $\dot M_{min}$, we set $\beta=0$.
For rates larger than $\dot M_{max}$ we use: 
$\beta
 = \dot M_{edd}/\dot M_{in}$

For NSs and BHs we chose:
$\dot M_{min}= 0.1 \times \dot M_{Eddington}$, and  
$\dot M_{max}= 10 \times \dot M_{Eddington}$. 
Although mass can likely be retained for even smaller
values of $\dot M_{min}$, our prescription 
leads to a conservative
estimate of mass gain and also focuses 
on intervals when mass transfer is most likely to 
produce high luminosities, and therefore to be detectable.
The upper limit reflects the fact that  
super-Eddington accretion 
has been observed in X-ray binaries: e.g., \citet{2017Sci...355..817I};
\citet{2014Natur.514..202B};
\citet{2016ApJ...831L..14F}.    
If we assume that the the accretion luminosity is 
$L_{acc} = 0.1 \times \dot M_{acc} c^2,$ then the infall rate onto
a NS or BH for
Eddington-limited 
accretion is $\approx M \times  2.4 \times 10^{-8} M_\odot$~yr$^{-1}$, where we
use $L_{Edd}\approx 1.3\times 10^{38}$~erg~s$^{-1}$ and where $M$ is the mass of the accretor.

For WDs, there
is a narrow range of infall rates for which mass can undergo nuclear
burning as it accretes, and can therefore be retained. [See, e.g.,
\citet{Iben.1982}; \citet{Nomoto.1982}; \citet{Shen.2007}.] The 
upper and lower bounds of this range depend on the value of the WD mass,
but at high masses, $\dot M^{\rm burn}_{min}$ may be a few times 
$10^{-7} M_\odot$~yr$^{-1}$, and $\dot M_{max}$ is
$\sim 10^{-6} M_\odot$~yr$^{-1}$.  At very low rates of accretion,
classical novae occur over intervals that can be as long as 
${\cal O}(10^5)$~yrs, ejecting much of the matter accreted between
explosions. At rates just under $\dot M^{\rm burn}_{min}$, however,
much of the accreted mass can be retained, because nuclear burning 
occurs during recurrent novae. These repeat on intervals that can range from
months to decades,
 and are less energetic than novae, allowing
processed material to be retained.          
We therefore take $\dot M_{min}$ to be the minimum rate of accretion
consistent with recurrent novae. 

To provide points of reference we note that, for many of the systems we 
consider, infall rates of $10^{-8} M_\odot$~yr$^{-1}$ to 
$10^{-5} M_\odot$~yr$^{-1}$ are associated with mass retention.
Since not all of the mass lost by the donor falls toward the inner binary,
the rate of accretion is generally only a fraction of the donor's rate
of mass loss. 
Donors must therefore have the high mass loss rates
associated with either giants or massive main-sequence stars,
if the rate of mass infall to be adequate to 
lead to genuine mass gain by the 
compact objects comprising the inner binary. 
Other stars
can, however, influence the orbital angular momentum and time-to-merger
of the inner binary.

\subsection{The Effects of Mass Retention}
Mass retention plays three important roles. 

{\bf (1)} Mass retention decreases the
time to merger. Equation 1 shows that the time for the components of the
inner binary to merge is inversely proportional to the product 
$M_1 M_2 (M_1+M_2)$ Thus, increases in the masses of star~1 and/or
star~2 decrease the time to merger. The effect is actually more pronounced,
because the addition of mass to the inner binary can also significantly alter
the orbital separation, which (for fixed orbital angular momentum) 
is proportional to 
$[(M_1+M_2)/(M_1^2)\, (M_2^2)]$. This means that increases in the 
masses of the components decreases the value of their orbital separation,
even when the orbital angular momentum is constant. 
Equation~1 indicates that this plays an even more significant role in
decreasing the time to merger.

{\bf (2)} Mass retention can change the physical nature of the
accretor. A carbon-oxygen (CO) WD, typically with mass 
below $\sim 1.15\, M_\odot$,   
will explode as a Type Ia supernova if it achieves a critical mass, $M_c$.
This critical mass may be the Chandrasekhar mass, $M_{Ch}$, with a value of
$\sim 1.38\, M_\odot.$ The exact value of $M_{c}$ depends, however, on detailed composition and
also
on the WD's spin. 
%A fast-spinning WD may have a much higher critical mass. If so,
%and if its final mass is greater than $M_{Ch}$ but smaller than its instantaneous
%critical mass, it may need to spin down before exploding.  The spin-down times
%may vary. Furthermore, even without spin down, there is a simmering time of
%about 1000 years from the time the critical mass is achieved to the time the explosion
%occurs. The upshot is that a CO WD may need to gain only about $0.4\, M_\odot$ to explode,
%and both it and its companion may continue to accrete after the WD has achieved $M_{Ch}.$
A slightly more massive WD, an oxygen-neon-magnesium(O-N-Mg) WD, 
will collapse to
become a NS after achieving its critical mass. 
Such a WD may require less than 
$0.2\, M_\odot$ to collapse.  
Similarly, a NS may collapse to become a BH if its mass exceeds a certain critical value.
That value is not yet well determined. To be specific we will take the upper
limit of the NS mass to be $2.2\, M_\odot$, which will also be the lower
limit of the BH mass in our simulations \citet{2017ApJ...850L..19M}, 
\citet{2011ApJ...741..103F}.   

{\bf (3)} Even if the nature of the accretor is unaltered,
mass retention changes the mass of the binary's components from the
values they would have achieved through single-star or binary evolution.   
BHs can undergo the largest mass increases possible among accretors 
that retain the
same physical natures.
This can happen when star 3 is itself a very massive star that 
could donate
a significant fraction of its mass to the close binary. 
Whatever the nature(s) of the 
accretors, mass added by star~3 alters the mass we measure
at the time of merger. Thus, if mass transfer is common, the masses
of the merging compact stars have a good chance of 
having been significantly changed between
the time the close binary was formed and the time its components merged.

\subsection{Flow of Angular Momentum}

%Mass leaving Star~3 carries angular momentum drawn
%from the angular momentum of the outer orbit:
%$L_{out}\sim M_3 (M_1+M_2)\, \sqrt{a_{out}}/\sqrt{M_1+M_2+M_3}$.

Although three-body motion can be complex, we will
focus on intervals during which the inner and outer orbits
are each well defined. The two compact objects occupy the inner orbit,
and the much larger outer orbit is defined by star~3 in orbit with the 
binary's center of mass. The orbital angular momenta are
$L_{in}=M_1 M_2 \sqrt{a_{in}/M_T}$, where $M_T=M_1+M_2$, and
$L_{out} = M_3 M_T \sqrt{a_{out}/M_{tot}}$, with $M_{tot} = M_3+M_T.$
The total orbital angular momentum is $\vec L=\vec L_{in}+\vec L_{out}$       

Gravitational radiation drains angular momentum from each orbit. 
Mass flowing from the system also carries angular momentum. Thus the
net flow of angular momentum from the 3-body system is negative, with the
ejection of mass potentially removing more angular momentum per unit time
than does gravitational radiation. If some of the exiting angular momentum
is drawn from the inner orbit, then the time  to merger decreases.

The flow of angular momentum within the system depends on a variety of
factors.  In many respects the mass flow configurations should be
similar to those observed in systems in which a
single compact object accretes matter from a  companion.
The rotating dipole 
component of the gravitational potential should play a significant role 
only when the incoming mass approaches the
inner binary. 
It is therefore likely that, in many cases, an accretion disk will be formed 
and that, as is found in for supermassive BH-BH binaries, the
inner binary clears a region just around it. In this case, the
less massive compact object will come closer to the edge of the 
circumbinary disk; a minidisk may then form around it.
This mode of accretion would eventually mean the components
of the inner binary could come to have nearly equal masses, even if they
had started with very different masses.

A circumbinary disk can play an active role in angular 
momentum transport and loss.
In the calculations 
performed for this paper, we did not incorporate direct effects 
produced by such a disk. We note, however, that both tidal
interactions between the disk and binary, and the release of even small 
amounts of mass from the outer disk, may tend to shrink the inner 
 binary.      

In addition to situations in which there is a circumbinary disk, 
it is likely that there are cases in which mass falls almost radially inward toward the inner binary's center of mass. In this case, the more massive component may be more likely to be the first to capture incoming mass. Alternatively, the mass flow may be well modeled by considering accretion from a dense medium within which the inner binary moves as it orbits its distance companion. Furthermore, if
one or both accretors are NSs or WDs, magnetic effects may play important roles in channeling mass toward them; or else they could produce a propeller effect, shooting mass from the system.
In addition, 
the compact objects can act as sinks of angular momentum when accreted
mass spins them up, or else as sources of angular momentum, if ejected mass
spins them down. 
In our evolutionary calculations we will 
not consider spin.

The considerations above indicate that the flow of angular momentum
may be complicated and that different processes may play dominant roles in 
different systems.  Nevertheless, there are important commonalities
that we try to capture.  First, of course is that angular 
momentum is carried by ejected mass. Second, that a good portion of the
angular momentum carried away is drawn from the orbits. Since the
orbital angular momentum of the outer orbit is generally much larger than that
of the inner orbit, most of the angular momentum carried away will 
be at the expense of the outer orbit. Nevertheless, even a small 
decrease in angular momentum of the inner orbit can decrease its
time to merger.    
In our calculations mass ejected from the vicinity of one of the three stellar 
components carries
away an amount of angular momentum that is proportional to the
 specific angular momentum of that star.

\subsection{Mechanisms for Mass Transfer and Internal
Loss of Angular Momentum} 

\subsubsection{Winds}

To model $\dot M_{wind}$, the rate of mass loss due to winds,
we implement an approach well suited to donors which leave WD
remnants, using a
version of the Reimer's wind which we have modified so that the
envelope of the star is exhausted when the core mass of $M_3$
has reached its final value (i.e., when star~3 has evolved to become
a WD). To compute the final mass of a WD-producing star, we use an
observationally established  
initial-mass/final-mass 
relationship.

This formalism allows the instantaneous value of the stellar mass,
$M_\ast(t)$, 
to be expressed in terms of the instantaneous value of the star's
core mass, $C(t)$, the initial value $M_\ast(0)$ of the stellar mass,
and a parameter $C_0,$ which is set to $0.2\, M_\odot.$  
\begin{equation} 
M_\ast(t) = \Bigg[M_\ast(0)^2 + 
\frac{\Big(C_{max}^2-M_\ast(0)^2\Big)\, 
     \Big(C^5-C_0^5\Big)}
{\Big(C_{max}^5-C_0^5\Big)}\Bigg]^\frac{1}{2} 
\end{equation} 
Wind mass loss in this model increases dramatically toward the
end of the giant's life, consistent with observations.      

The stellar wind, $\dot M_\ast(t)$ is just the time derivative of the mass;
this includes terms involving the $\dot C(t)$, which is proportional to the
stellar luminosity.  
Because the donor is a giant, releasing mass in many directions, 
only a fraction $f$  of the ejected mass can be captured 
by the inner binary. 
We use the following formula.
\begin{equation} 
\gamma = \kappa\, \Bigg(\frac{R_3}{R_L}\Bigg)^{\Big(2-\frac{R_3}{R_L}\Big)}, 
\end{equation} 
where $R_3$ and $R_L$ are the instantaneous values of the donor's
physical radius and Roche-lobe radius, respectively, and $\kappa$ is
a constant, whose value we have taken to be $0.5$ in the calculations 
described here.  
The functional form above ensures that the rate of wind capture is
roughly equal to $\kappa$ when the donor fills or nearly fills its
Roche lobe. Because $R_L$ is proportional to $a_{out}$, the capture
fraction falls off as $1/{distance}$ for systems close to Roche-lobe filling.
On the other hand, gravitational focusing plays a smaller role 
at larger separations, so the capture fraction should fall off as 
roughly $(1/{distance})^2$.
The exponential ensures that this is the case at larger distances.

The rate of mass infall to the binary is $\gamma\, \dot M_3.$
As we will show in \S 3, winds alone can produce important
changes in the inner binary, decreasing its time to merger and
sometimes transforming the nature of
its components. 

\subsubsection{Roche-lobe-filling}

The donor star expands with age so that, if the initial size of the 
outer orbit, $a_{out}(0)$, is small enough, the donor may come to fill its
Roche lobe.  
Once the Roche lobe is filled, there will either be an instability that
leads to a common envelope (see below) or else there can be a relatively
long epoch during which mass loss from the donor increases,    
with a fraction of its mass channeled through the region around the
L1 point.  
When the donor
star is a subgiant or giant, this epoch ends when the stellar envelope is
exhausted by the combination of mass loss and the growth of the stellar
core.

\subsubsection{Common envelope}

When a star in a binary fills its Roche lobe, the transfer of mass to 
its companion
will change the dimensions of the Roche lobe. At the same time,
the loss of mass from the donor may
 alter its radius. If the Roche lobe shrinks at a rate faster than
the star can adjust, mass transfer will proceed on a dynamical 
time scale and a 
common envelope will encompass both the core of the 
donor and the accretor \citet{1977ApJ...215..851W, 1976IAUS...73...75P}.

Generally, the envelope will be ejected over an interval
 of $10^4-10^5$ years. Unless super-Eddington accretion occurs during
this short-lived phase (a process that is invoked to form
double NSs, for example), 
little or no mass may be gained by the donor's companion.
The common envelop can, however, have a pronounced effect
on the orbital angular momentum
 of a binary \citet{2005MNRAS.356..753N}. By imparting
angular momentum to mass in the envelop, the components
 of the binary spiral closer
to each other while ejecting the envelope. Angular momentum
considerations can be used to express the 
final orbital separation in terms
 of the
initial system parameters. 

When the accretor is a compact binary, it
 too can impart angular momentum to the 
common envelope. This helps the envelope to escape, 
while at the same time bringing the
binary's components closer
 to each other. The question of how much angular momentum is
lost by the inner binary is difficult
 to answer, if only because
the analogous question has 
proved challenging even for the simpler two-body systems.  
A formalism well suited
 to computing the effects of the common envelope on
the separation of the inner binary  focuses on the role of angular
momentum.

\begin{equation} 
\frac{\Delta\, L}{L} = g\,  \frac{\Delta\, M}{M}.   
\end{equation}
The value of $g$ is uncertain. Nelemans and Tout (2005) estimated that its value
is $\approx 1.6$ for the common envelope phase that produced  a set of
double WDs.

\subsection{Detectability} 

Mass transfer is potentially detectable.   
During the epoch in which the accretors are able to retain mass,
their X-ray luminosities, in our model (\S 2.4) are
within a factor of 10 of the Eddington luminosity. Many of these systems
would have X-ray luminosity above $10^{38}$~erg~s$^{-1}$ or
$10^{39}$~erg~s$^{-1}$
during this interval and would be detectable even in external galaxies. The 
slower accretion that would take place over longer times prior to the
high-accretion phase would be dimmer, but nevertheless 
detectable in the Milky Way, Magellanic
Clouds and, during some intervals,
 even in M31. The X-ray emission could also show signs of a 
short-period component, due to the motion of the inner binary. The inner
orbital period
decreases as mass transfer proceeds. For nearly edge-on 
orientations, emission from the active accretor(s) 
could be lensed, producing a distinctive
periodic signature. In most cases the donor star would be a giant, and the
system would be identified as a symbiotic binary. Any short-period
signature associated with the motion of the inner binary would be the
tip-off that the accreting system is a binary.  
The duration of interval when the X-ray emission is detectable depends on
the flux (hence the distance to the source), and on the
lifetime of high-wind phase of the donor, which is longer for less
massive donors.

\subsection{Calculation of the Evolution}

Our calculations start
at the time when the
core mass of the donor, star~3, is 
$0.2\, M_\odot$, and continue until the donor's
envelope is exhausted.
We increment the core mass of star~3 by $dc$, 
compute the time $dt$ 
it would take for the core to have grown by this amount, 
and determine the star's mass, radius,
 and rate of wind mass loss at the new time. 

The donor ejects mass from the system at the rate: 
$(1-\gamma)\, \dot M_{winds}$.
The ejected mass carries 
 specific orbital angular equal to $v_3$ times the orbital angular   
momentum of star~3. $v_3$ is one of the model's adjustable parameters.
The ejected angular momentum comes entirely at the
expense of $L_{out},$ the angular momentum of the outer 
orbit. The remainder of the mass, $\gamma\, \dot M_{winds}$  
flows toward the less massive star, star~2. We use the considerations
described in \S 2.3 to compute $\beta_2,$ and 
consider that the rest of the mass,
$(1-\beta_2)\,\gamma \, \dot M_{winds}$, is incident on star~1. 
We compute $\beta_1$ and assume that any 
mass that cannot be retained
by star~1 is ejected from the system, 
carrying $v_1$ times the specific angular momentum of star~1;
$v_1$ is another adjustable parameter.  Because star~1 is part of
both the inner and outer binary, we have to subtract angular momentum from
both $L_{in}$ and $L_{out}$. To do this we subtract from $L_{in}$ ($L_{out}$) 
the (rate of mass loss from the inner binary)  
multiplied by $v_1$ times the (specific angular momentum associated
with the inner [outer] orbit of star~1). 

Independently draining angular momentum from the inner and outer orbits
is well suited to cases in which  
the orbits are orthogonal. It is also appropriate for any 
system without a direct link between the orbital angular 
momentum of the inner and outer orbits: for example, when
accretion onto the compact objects is approximately spherical or else
when angular momentum is dissipated within an accretion disk.  
We discuss cases in which there is a link between the inner and outer orbits
in \S 4. 

Note that the above approach paints the evolution with a broad brush.  It 
includes the key processes determining the fates of these hierarchical
triples, but does not attempt to track these processes in detail.
In fact, the true physical processes are complex and not yet well understood.
For example, the rate of mass loss due to winds in evolving stars
is not likely to be steady, and both first principles calculations and 
inferences from observations are challenging, especially for
massive stars and  for stars in the end stages of stellar evolution.
The focusing of winds is suggested by observations, but exactly how
this depends on the mass loss rate, the speed of exiting mass, and 
irradiation from the accretors still needs to be understood. There are
also significant uncertainties about the infall of mass to the 
inner binary, the ability of this mass to reach the
components of this binary,
and the ability of these components to retain matter. Finally, an important
question is: how much angular momentum is carried by matter exiting the system?

Any calculations that attempt to model all of these processes would have to 
include the above-mentioned significant and difficult-to-quantify uncertainties.
Our approach captures the important features of the evolution;and
the extent of the associated uncertainties can be gauged by conducting
a range of simulations with different values of a small number of
input parameters.   
As we will see, the character of the results depend on only a few key 
assumptions.

\section{Results}

\subsection{Individual Systems} 
\begin{figure}
        \includegraphics[width=\columnwidth]{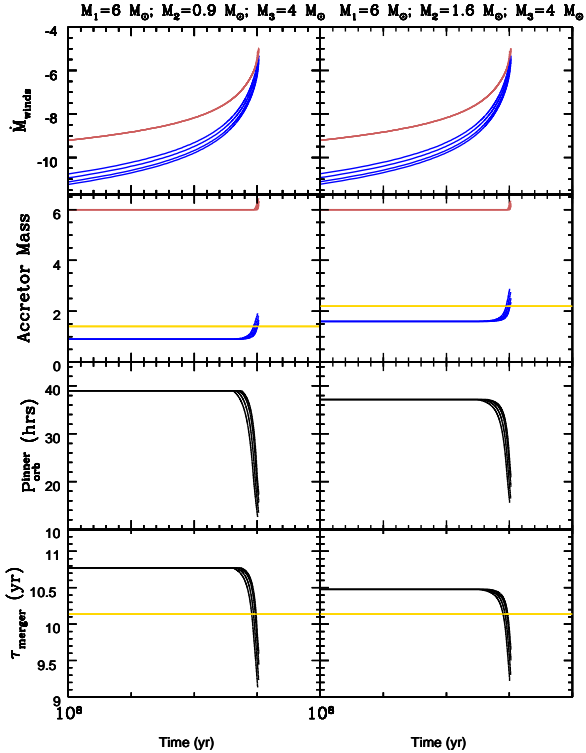}
    \caption{{\bf Sample evolutions:} In both cases $M_1 = 6\, M_\odot$ and 
$M_3=4\, M_\odot$. Left: $M_2 = 0.9\, M_\odot.$ Right: $M_2 = 1.8\, M_\odot.$
The top panels show $\dot M_{wind}$ (red), and $\dot M_{in},$ the rate of mass
infall to the binary (set of dark blue curves). The next panel down shows the
values of $M_2$ (blue, lower), and $M_1$ (red, higher). The straight gold line on the left marks the $M_{Ch}$ and the gold line on the right corresponds to a 
maximum NS mass of $2.2\, M_\odot.$ Proceeding downward, the next
panel shows the orbital period of the inner binary. The final panel
shows the ratio of the final value of $\tau_{merge}$ to 
$|t-t_{merge,initial}|$ versus $M_3$ (left) and $a_{out}(0)$ (right). 
}
    \label{fig:example_figure}
\end{figure}
\begin{figure}
        \includegraphics[width=\columnwidth]{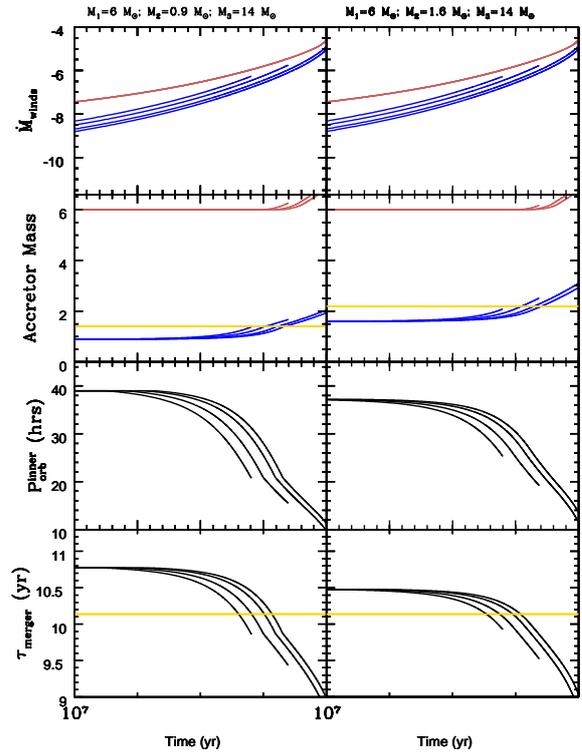}
    \caption{The same as Figure 1, except that the donor masses on both the
left and right are $14\, M_\odot.$ All other system parameters are
identical.
        }
    \label{fig:example_figure}
\end{figure}

In Figures 1 and 2 we show the results of evolving 16 individual systems.
Figure 1 shows, side by side, two sets of evolutions, each starting
with a single value of $M_1$ ($6\, M_\odot$; corresponding to a stellar-mass
BH) and $M_3$ ($4\, M_\odot$; corresponding to a donor star of
relatively modest mass). 
On the left, star~2 is a $0.9\, M_\odot$ CO-WD,
and on the right it is a $1.6\, M_\odot$ NS.
For both the WD and NS, we show the evolutions of $4$ systems which
differ from each other in the initial radius of the outer orbit, which
ranges from $10$~AU to $23$~AU. 
 For all
of the initial orbital separations, the WD (left) achieves 
the Chandrasekhar
mass, exploding as an SN~Ia.  Similarly, for all the initial 
separations considered,
 the NS (right) reaches the critical mass and undergoes an AIC to become a BH.
  In all cases 
the time-to-merger decreases from above the Hubble time
to a value on the order of a few billion years. The orbital period of the inner
orbit also significantly decreases. In all cases the time interval during which the
most significant changes occurred lasts for a few times $10^6$ years.   
Only for the two smallest initial values of
$a_{out}$ does star~3 fill its Roche lobe. 
Roche-lobe filling leads to a slightly larger accretor final mass 
(a few tenths of a solar mass). The results for wider separations, where the 
donor never fills its Roche lobe, show that significant  
changes can be effected in the inner binary through the agency of winds alone. 

From the perspective of the processes at work, the key item of note is that,
even though the donor star is a subgiant or giant throughout the time
shown, there are essentially no changes in the properties of the inner orbits
until the rate of wind mass loss is high. This is because, in our
models, the accretors gain mass only for large infall rates. The
infall rate depends on the donor's mass loss rate, and also on the size of the
outer orbit relative to the donor's radius. 
The donor's winds and radius increase with time. 
In addition to winds, Roche-lobe
filling provides an effective way for the inner binary to gain mass. 
Even for Roche-lobe-filling systems, winds are important 
because an epoch of heavy winds precedes 
Roche-lobe filling. Thus, there is not a dramatic change at the time
of Roche-lobe filling unless the donor is so massive that a common envelope forms.
 
Figure 2 differs from Figure 1 only in the mass of the outer star, which
was taken to be $14\, M_\odot$ for both the WD (left) and NS (right) cases.
The same set of $4$ orbital separations were chosen.  In this case, 
two of the evolutions terminate at relatively early times. 
These correspond to systems in which star~3 fills its Roche lobe 
when it is more massive than the inner binary, so that a 
common envelope forms. During the very short duration of the
common envelope, angular momentum continues to be lost by the inner orbit,
but (in our model) no mass is gained by the compact objects in the inner orbit.  

The most obvious difference between the cases with
$M_3=14\, M_\odot$ (instead of $4\, M_\odot$, as considered in Figure 1)
is the availability of more mass\footnote{Note, however, that in the evolutions shown,
the availability of additional mass did not lead to significantly larger
increases in the masses of the accretors. This is because of the interplay in our model
between the rate of mass infall and the ability of the accretors to accept
and retain mass.}. 
There is, however,
 another difference that plays
an important role: the more massive star evolves on a 
shorter time scale, so that
the transitions take place over shorter times.

\subsection{Large Numbers of Systems}  

\begin{figure}
        \includegraphics[width=\columnwidth]{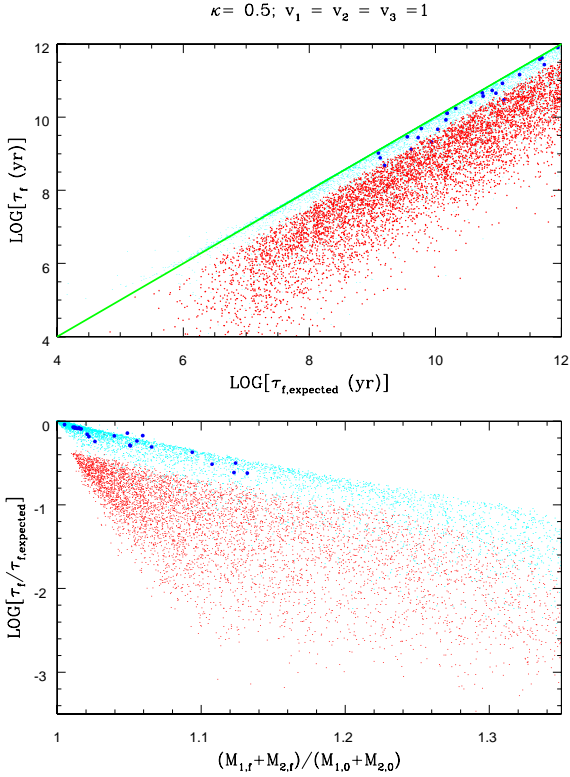}
    \caption{Results of calculations for 100,000 hierarchical triples; 
the model parameters are given in the figure's top label.  
{\sl Top panel}: the logarithm to the base ten of the final time to merger 
($\tau_f$,
the time to merger as measured after mass transfer is finished) is plotted
against the the logarithm to the base ten of $\tau_{f,expected}$,
 the time to merger that would have 
been expected, had no mass transfer occurred. 
{\sl Bottom panel}: The logarithm to the base 
ten of the ratio $\tau_f/\tau_{f,expected}$ is plotted 
versus the ratio of the final total mass of the inner binary to 
its initial total mass. 
Points in cyan (lightest points)
 are systems 
experiencing only wind mass transfer; larger and somewhat darker (blue) points are systems that
have experienced wind mass transfer and then stable mass transfer while the
donor fills its Roche lobe; darkest points (red) experienced wind mass transfer
and then a common envelope after Roche-lobe filling. 
}
    \label{fig:example_figure}
\end{figure}
\begin{figure}
        \includegraphics[width=\columnwidth]{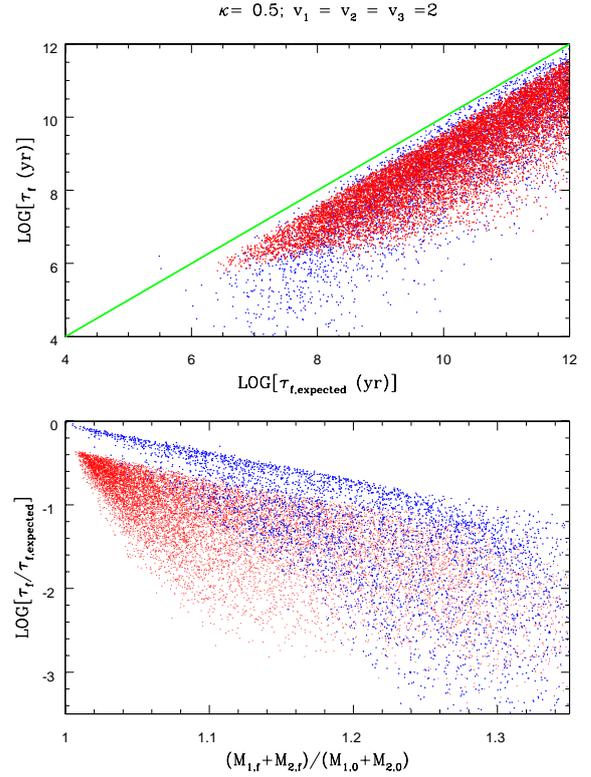}
    \caption{
Same set-up as for Figure 3, with simulation parameters given
in the top label.  
Color (gray scale) coding of points is
described in \S 3.2.1 and also in the caption to Figure 3.}  
    \label{fig:example_figure}
\end{figure}
\begin{figure}
        \includegraphics[width=\columnwidth]{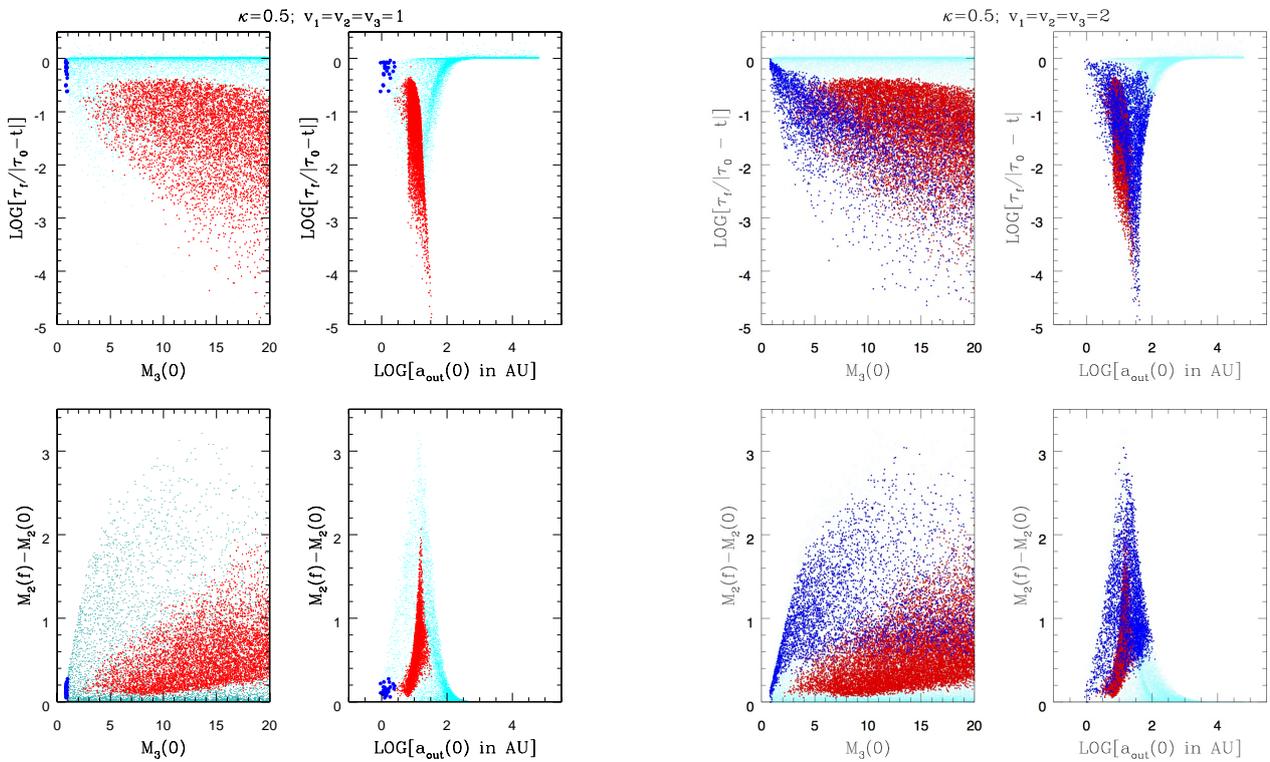}
    \caption{Results of the same calculation shown in Figure~3.
The change in $M_2$ is shown along the vertical axes in the bottom panels. 
{\sl Right:}~$M_3$
is plotted along the horizontal axis; {\sl Left:}~$a_{out}(0)$ (in AU)  
is plotted along the horizontal axis. 
In the top panels the
quantity $\tau_f/|\tau_0-t|$ is plotted along the vertical axis.    
Color (gray scale) coding of points is
described in \S 3.2.1 and also in the caption to Figure 3.}  
    \label{fig:example_figure}
\end{figure}
\begin{figure}
        \includegraphics[width=\columnwidth]{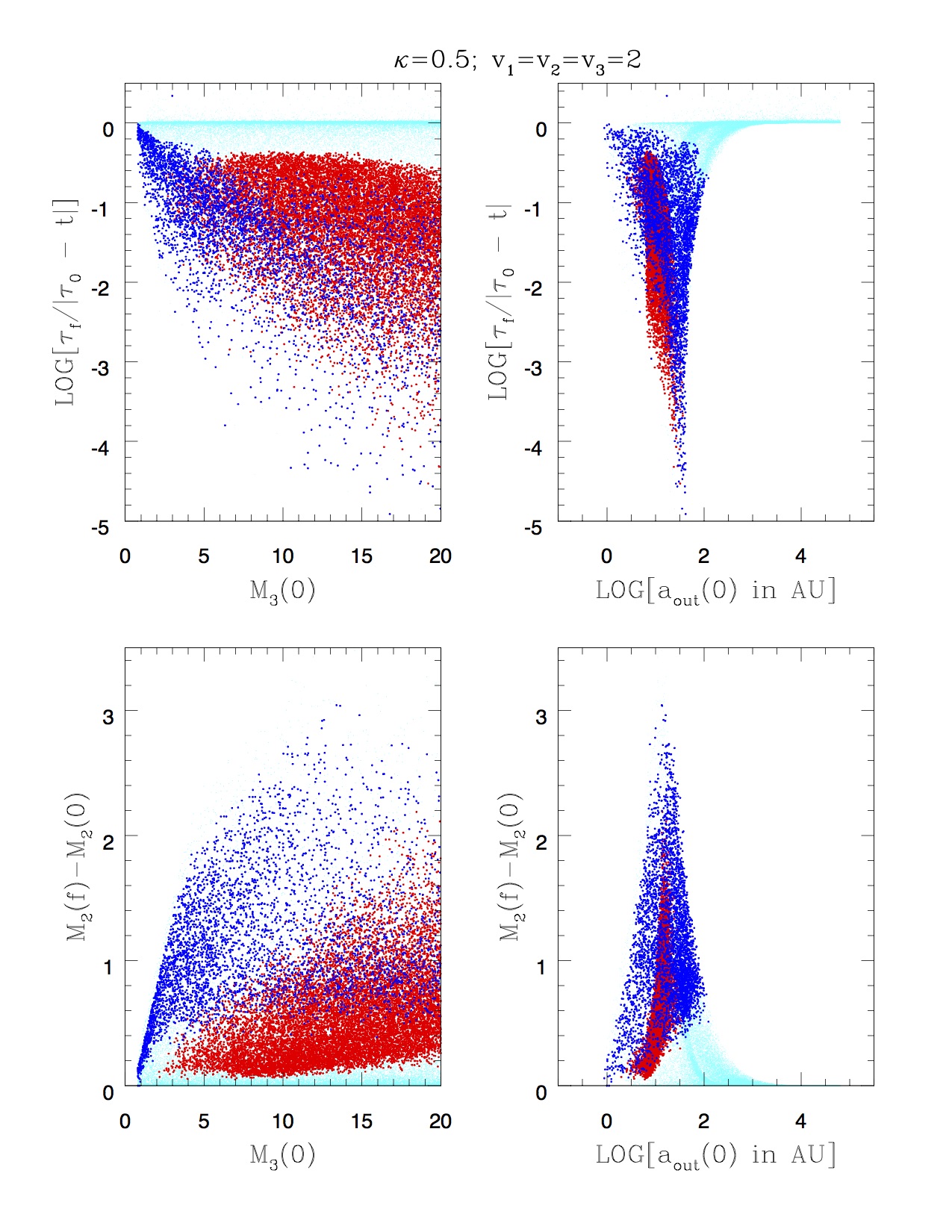}
    \caption{
Same as for Figure 5, with simulation parameters shown in the top label. 
Color (gray scale) coding of points is
described in \S 3.2.1 and also in the caption to Figure 3.}  
    \label{fig:example_figure}
\end{figure}

% model1:standard;
%model12 0 for dL_out, low-mass donor; model3: 1/2 for dL_out
\begin{table*}
        \centering
        \caption{
        Numbers of events per 33,333 triples}
        \label{tab:example_table1}
        \begin{tabular}{lccccccccccc} % 12 columns, alignment for each
                \hline
                \hline
                      &       &          &                       &  
	       & X-WD 
               & X-WD 
	       & X-WD 
	       & X-NS 
	       & NS-NS  
	       & WD-WD  	  
	       & WD-WD\\  	  
                $v_1$ & $v_3$ & $\kappa$ & $N_{< 0.5\, \tau(0)}$ & $N_{< 0.1\, \tau(0)}$  
               &  $\Downarrow$ 
               &   $\Downarrow$ 
               &  $\Downarrow$ 
               &  $\Downarrow$ 
               &  $\Downarrow$   
               &  $\Downarrow$   
               &  $\Downarrow$ \\
                      &       &          &                       &  
	       & X-Ia 
               & X-NS 
	       & X-BH 
	       & X-BH 
	       & BH-BH  
	       & NS-NS  	  
	       & Ia-Ia\\  	  
                \hline
                \hline
 % 0.00 &    0.00 &    0.50 &       5318 &       2826 &        620 &        361 &         93 &        889 &         98 &         18 &         36 \\
 % 0.50 &    0.50 &    0.50 &       5953 &       3143 &        723 &        428 &         65 &        998 &         86 &         23 &         27 \\
 % 1.00 &    1.00 &    0.35 &       6380 &       2834 &        689 &        499 &         32 &       1016 &         72 &         23 &         11 \\
 % 1.00 &    1.00 &    0.50 &       6913 &       3663 &        775 &        542 &         71 &       1206 &         84 &         16 &         21 \\
 % 1.00 &    1.00 &    0.75 &       7446 &       4449 &        903 &        541 &         70 &       1272 &        119 &         25 &         23 \\
 % 2.00 &    2.00 &    0.50 &       9118 &       4510 &        851 &        741 &         47 &       1357 &         78 &         30 &         10 \\
 % 2.00 &    0.00 &    0.50 &       5443 &       2898 &        641 &        451 &         33 &        914 &         64 &         19 &          9 \\ 
 % 0.00 &    2.00 &    0.50 &       8973 &       4394 &        875 &        696 &         91 &       1430 &        123 &         36 &         35 \\ 
% below is from vary_new 
   0.00 &    0.00 &    0.50 &       5282 &       2790 &        618 &        361 &         93 &        889 &         98 &         18 &         36 \\ 
   0.50 &    0.50 &    0.50 &       5911 &       3090 &        721 &        400 &         82 &       1000 &        102 &         21 &         36 \\ 
   1.00 &    1.00 &    0.35 &       6321 &       2742 &        709 &        475 &         52 &       1033 &         82 &         22 &         27 \\
   1.00 &    1.00 &    0.50 &       6853 &       3538 &        801 &        498 &        101 &       1217 &        109 &         16 &         37 \\
   1.00 &    1.00 &    0.75 &       7408 &       4368 &        929 &        490 &        108 &       1287 &        137 &         25 &         43 \\
   2.00 &    2.00 &    0.50 &       9024 &       4469 &        912 &        718 &         68 &       1419 &        110 &         30 &         20 \\
   2.00 &    0.00 &    0.50 &       5369 &       2798 &        662 &        389 &         77 &        930 &         82 &         22 &         28 \\
   0.00 &    2.00 &    0.50 &       8923 &       4344 &        873 &        693 &         91 &       1430 &        123 &         35 &         35 \\
               \hline
                \hline
        \end{tabular}
\end{table*}
To identify the sets of initial parameters (donor masses,
orbital separations) that lead to significant increases in the mass of the
inner binary and decreases in the time to merger, 
we conducted a set of simulations, 
each starting with tens of thousands of hierarchical triples.
To generate each hierarchical triple, we started by generating the value
of $M_1$, selecting uniformly between $0.6\, M_\odot$ (corresponding to 
a CO WD) and $7\, M_\odot$ [corresponding to 
a stellar-mass BH with mass typical of those discovered in nearby
X-ray binaries; \citet{2016A&A...587A..61C};
\citet{2006ARA&A..44...49R}].  
We allowed $M_2$ to have any mass smaller than $M_1$ but
larger than $0.6\, M_\odot$.
The value of $M_3$ was chosen to be in the range from $0.8\, M_\odot$ (roughly 
corresponding to 
the minimum mass of a star expected to evolve within a Hubble time) 
and $20\, M_\odot.$ 
The next step was to choose the time-to-merger for the inner binary: 
$\tau_{merge}$ was selected to be in the range between
($0.1\times$ the main-sequence lifetime of star~3) to ($10^{12}$ years),
with the exponent chosen from a uniform distribution. 
We then used Equation~1 to compute the radius of the inner
orbit. Equation~4 defines the minimum radius of the outer orbit. We
selected the maximum orbital radius to be as large as $10^4$ times the
maximum radius of star~3, selecting the exponent from a uniform 
distribution\footnote{With this formulation, many outer stars are in orbits so wide that they cannot
send a significant amount of mass to the neighborhood of the
inner  binary. Our goal, however, is to use  the calculations to explore
 the outer limits, beyond which mass transfer does not significantly change the inner binary.}.
We then computed the evolution of each individual system, ending at the time
when the envelope of the donor was exhausted.  
Each simulation was defined by the values of: 
$\kappa,$ $v_1, v_2,$ and  $v_3.$

%        sep_out_min=sep_in*y_min
%2         lsep_out=4.*ran1(idum)
%          sep_out=(10.**lsep_out)*rstar(m30,cmax)+sep_out_min

\subsubsection{Times to Merger and Mass Increases}

Figures 3 and 4 illustrate 
and quantify (1)~decreases in the time to merger 
and (2)~ mass gains by the inner binary.
Points in cyan (lightest color) correspond to systems in which the donor
never filled its Roche lobe; all mass transfer proceeded through winds.
Points in red (same size as cyan points, but darker) 
correspond to systems in which 
mass transfer occurred through winds, but the donor filled 
its Roche lobe at a time when
it was more massive than the binary. During the ensuing
common envelope phase, no additional
mass was gained by the binary, but the inner binary did lose orbital angular
momentum as it helped to eject the common envelope.   
Points in blue (dark and larger than the others) are systems in which the
donor filled its Roche lobe and was able to continue giving mass 
to the inner binary in a stable manner.  

In the top panel of Figure~3
 we consider the  time-to-merger as measured at the end of mass transfer, $\tau_f$. The logarithm to the base 10 of   $\tau_f$ is plotted along the vertical axis. Along the horizontal axis is the logarithm of the ``expected'' time-to-merger.  The definition of $\tau_{f,expected}$ is: 
the original value of the time to merger (i.e., as calculated 
prior to mass transfer), minus the time duration of mass transfer. 
Thus, the value of $\tau_{f,expected}$ would have been the remaining 
time to merger had no mass transfer occurred.  
The diagonal green line corresponds to the case in which mass transfer
has no effect on the time to merger.
The factor by which the time-to-merger decreases is often as small
as
$0.01$, with some systems exhibiting even more dramatic decreases.   

The bottom panel explores the relationship between the decrease in 
the time to merger and the increase in the masses of the components 
of the inner binary. Along the horizontal axis is plotted the ratio 
of the total mass of the inner binary after mass transfer, 
to its value prior to mass transfer. Along the vertical axis is 
the logarithm to the base 10 of $\tau_f/\tau_{f,expected}$. 
The panel shows that the time to merger can decrease significantly, 
even if the inner components of the binary gain very little mass. 
As expected, common envelope evolution can lead to significant 
shortening of the time to merger while the mass of the inner binary 
experiences only a marginal increase. On the other hand, 
Roche lobe filling and even winds alone or winds followed 
by a common envelope phase, can lead to both significant 
mass increases and to significant decreases in the time to merger.

The same quantities are plotted in Figure~4, but the parameters
that control angular momentum loss are larger ($v_1=v_2=v_3=2$), 
indicating that
more angular momentum is carried away by mass leaving the system.
In particular, the outer orbit is more likely to shrink, so that
larger numbers of donors will fill their Roche lobes. A larger fraction
of    
common-envelope and Roche-lobe-filling systems inhabit the panels of 
Figure~4. As a consequence, typical times to merger decrease even more
than in Figure~3, and the fractional change in mass of the inner binaries is
larger. Two messages emerging from these simulations are the following.
{\bf (1) There are significant effects even with low angular momentum loss.
(2) Increasing the amount of angular momentum lost 
increases the magnitude of the effects and the numbers of systems experiencing them. 
}

\subsubsection{What characteristics of the outer binary produce the most pronounced changes?}

Figures~5 and 6 relate the changes in 
the inner binary to the initial characteristics of the outer binary.
In the right-hand panels we consider the effect of varying $a_{out}(0)$, 
the initial radius of the outer orbit. By this we mean  its radius just prior to
the epoch of mass transfer from star~3. 
In the left-hand panels we explore the influence of the value of 
$M_3(0)$, the initial mass of the donor.

The bottom panels 
plot $(M_2(f)-M_2(0))$, the
change in the mass of star that was initially the least massive stellar
remnant in the inner binary.   
The top panels plot 
the quantity $\tau_f/|\tau_0-t|$
\footnote{This functional form yields values almost
equal to the ratio $\tau_f/\tau_0$ when the evolutionary time is shorter
than the initial time to merger, but gives large values when the
merger time is already so short that the binary may merge event before star~3
is fully evolved.
Such systems are interesting because there would be mass in the vicinity of
the binary as it merges, potentially producing detectable 
electromagnetic signatures to accompany the
emission of gravity waves. From the perspective of altering the time
to merger, however, the effects are likely to be slight and we therefore 
designed the functional form $\tau_f/|\tau_0-t|$ so that small values
would allow us to easily identify the systems in which the time-to-merger
was most altered by mass transfer.}.  

Figures 5 and 6 share several common features.
First, with regard to the influence of the donor mass: while more massive donors
can increase the secondary's mass by the most, something just over $3\, M_\odot$
for $M_3(0) = 20\, M_\odot$, even low mass donors can be
responsible for significant increases. A star of $2\, M_\odot$
($5\, M_\odot$)  can increase the 
secondary mass by nearly $1\, M_\odot$ ($2\, M_\odot$).
Furthermore, these large changes can be made through the action of winds.
(This is more obvious in the left-hand panels, which include fewer of the
somewhat larger points associated with Roche-lob filling.)

Another commonality is that the initial orbital separation makes a
big difference to both the inner-binary mass increase and the 
decrease in the time to merger.
Furthermore,  there is a range of values of $a_{out}(0)$     
over which the changes in mass and merger times are sharply peaked.
For $v_3=1,$ the peak lies in the range between about $10$~AU and $20$~AU. 
This peak moves out to  larger values of $a_{out}(0)$  
when there is more angular momentum loss. This is because the loss of 
angular momentum from the outer orbit decreases the value of $a_{out};$
thus, the initial value of $a_{out}$ is much larger than the values at which
most of the mass transfer occurs.

\subsubsection{Comparing results across simulations}

Table 1 shows the results for a set of 8 simulations.   
The first two columns show the values used for $v_1$ and $v_3$.
These are the constants of proportionality between the angular momentum
per unit time carried away from the outer orbit
 by mass exiting from star~1 and star~3,
respectively, and the specific angular momentum of these stars.
Matter incident on the binary first travels to $M_2$. Any mass that cannot be accreted by $M_2$ then travels to $M_1,$ and mass that cannot be accreted by $M_1$ exits the system, Since matter does not exit directly from $M_2$, the value of $v_2$ 
doesn't
directly influence the evolution. 

The model parameter
$\kappa$ appears in the third column. The fourth column shows the number of systems, $N_{< 0.5\, \tau(0)}$, for which 
the times to merger were reduced by more than a factor of $2$. 
$N_{< 0.5\, \tau(0)}$ is a rough estimate of the numbers of systems in
each simulation that are significantly influenced by mass flowing from the
outer star toward the inner binary; its value ranges from 
$\sim 5300$ to $\sim 9000$,
representing between $5\%$ and almost $10\%$ 
of the initial set of triples. Because most members of the initial set
did not start with values of $a_{out}$ small enough to allow 
significant mass flow to the inner binary, the value of $N_{< 0.5\, \tau(0)}$
provides a guide to the numbers of sets of initial conditions that lead to 
significant changes.    

There are some clear trends in the values of $N_{< 0.5\, \tau(0)}$. 
For example, larger changes are effected
for larger  
values of $\kappa$. The third, fourth, and fifth rows are for systems with
$v_1=v_3=1$, but for $\kappa$ equal to $0.35$, $0.50$, and $0.75$, respectively.
The increase in the effectiveness of mass transfer
with increasing $\kappa$ is expected, unless incoming mass is processed 
much less efficiently by the inner binary when the rate of
mass infall is large.

Second, more angular momentum loss, particularly from the outer orbit,
increases the numbers of systems experiencing significant effects. 
This can be seen by comparing the first, fourth, sixth, and eighth rows.
The more angular momentum lost from the outer orbit, the larger the numbers of
outer stars that can be pulled in 
close enough to the inner binary to donate significant amounts of mass.

The fifth column shows 
the number of systems, $N_{< 0.1\, \tau(0)}$, for which     
the times to merger were reduced by more than a factor of $10$.
We find that, across simulations, 
$N_{< 0.1\, \tau(0)} \approx 0.5\, N_{< 0.5\, \tau(0)}$. This includes
binaries which started having times-to-merger larger than $\tau_H.$

The sum of the sixth and seventh columns provides the numbers of WD-containing
close binaries in which a WD made a transition to an SN~Ia or to an NS,
while the nature of the companion stayed the same.
The numbers of transitions to NSs are smaller than the numbers of SNe-Ia
in these simulations because the initial WDs masses needed to transition to a
NS span only the narrow range between $1.15\, M_\odot$ and $1.38\, M_\odot,$
while SNe~Ias occur in our simulations from $0.6\, M_\odot$ to $1.15\, M_\odot$. 
Although the ratio of the lengths of the starting ranges is only $0.41$, the 
ratio of AICs to SNe~Ia is larger because less mass is generally 
needed to effect an AIC. 

The sum of the sixth and seventh columns is roughly equal to 
$0.3\, N_{< 0.1\, \tau(0)}$. The numbers of NSs that make transitions to BHs,
while their companion does not transition is also roughly equal to
$0.3\, N_{< 0.1\, \tau(0)}$. The numbers of binaries that experience double transitions (e.g., NS/NS to BH/BH) is roughly $10\%$ the number of single transitions
(e.g., NS/X to BH/X). 

The upshot of these comparisons is that, although the amount of mass channeled 
to the
inner binary, and the loss of orbital angular momentum from the outer binary, 
both 
play significant roles, changes in the time-to-merger and 
transitions of compact objects should both be common.        

\subsubsection{X-Ray Hierarchical Triples}

Mass approaching close to or
accreting onto one of the compact objects is expected to emit X-rays. For each particle,
the accretion luminosity
is typically a significant fraction of its rest mass.  
Compact binaries receiving  mass emitted by winds or through Roche-lobe-filling can therefore be
very bright. With intrinsic and/or lensed 
luminosities that may be near or above $10^{40}$ erg~s$^{-1}$, 
they would be detectable in external galaxies. 

An interesting feature of these systems is that, although they would appear to be
X-ray binaries, they are actually 
X-ray hierarchical triples. They could exhibit periodic or quasiperiodic signatures
related to the orbital period of the inner binary. At the same time, they would
exhibit features characteristic of symbiotic binaries.
 
If hierarchical triples are common, a  
significant fraction of bright variable X-ray sources are likely to actually
 be {\sl accreting binaries} 
with wide-orbit companions. Archival studies searching for variability among
bright X-ray sources, and also identifying counterparts across wavebands
could identify these systems.
The population would consist of pre-merger binaries, but also many others which cannot
merge in a Hubble time. In addition, some short-orbital period binaries of many types might
be accreting from low-mass companions. An example of a system in which one of the 
components of the inner binary may not be a stellar remnant is a cataclysmic variable, in 
which the companion may be a very \-low mass star, perhaps itself degenerate.
The orbital period would be on the order of a few hours.  
It is important to conduct archival searches for X-ray triples. 
 
\begin{figure}
        \includegraphics[width=\columnwidth]{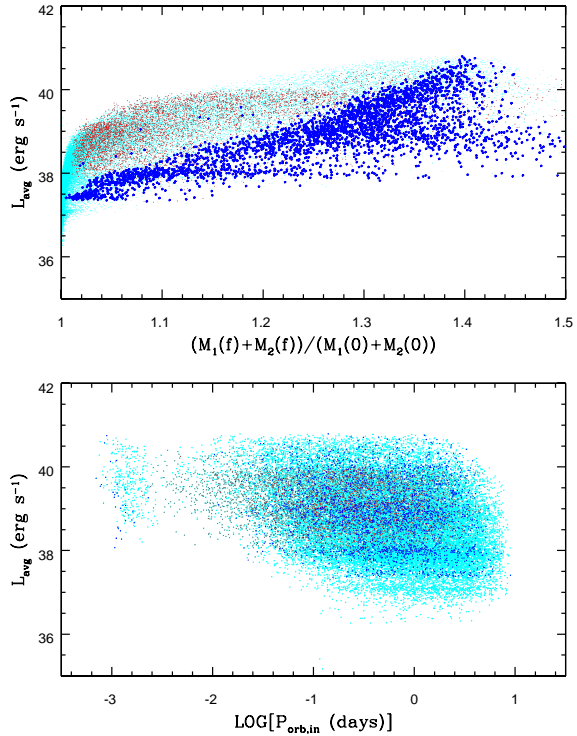}
    \caption{
{\bf Average X-ray luminosity of the least massive accretor versus
the ratio of the inner binary's final mass to its initial mass 
(upper panel) and final orbital period (lower panel).} As in the previous
figures, systems in: 
{\color {Cyan}} 
transferred mass entirely through winds; 
{\color {red}} 
transferred mass entirely through winds and then produced a
common envelope when the donor filled its Roche lobe; 
{\color {Blue}} 
transferred mass through winds and then through the L1 point as well,
during Roche-lobe filling. 
The luminosity of $M_2$ is computed in each time step and ts average
value over the time during which it is larger than 
$10^{32}$~erg~s$^{-1}$ is computed.  
 }
    \label{fig:example_figure}
\end{figure}
   
\begin{figure*}
%        \includegraphics[width=\columnwidth]{/home/rd/mnras/picture.pdf}
%        \includegraphics[width=11 in]{/home/rd/mnras/picture.pdf}
	\hspace{-1.6in} 
	\vspace{-3in} 
         \includegraphics[scale=1]{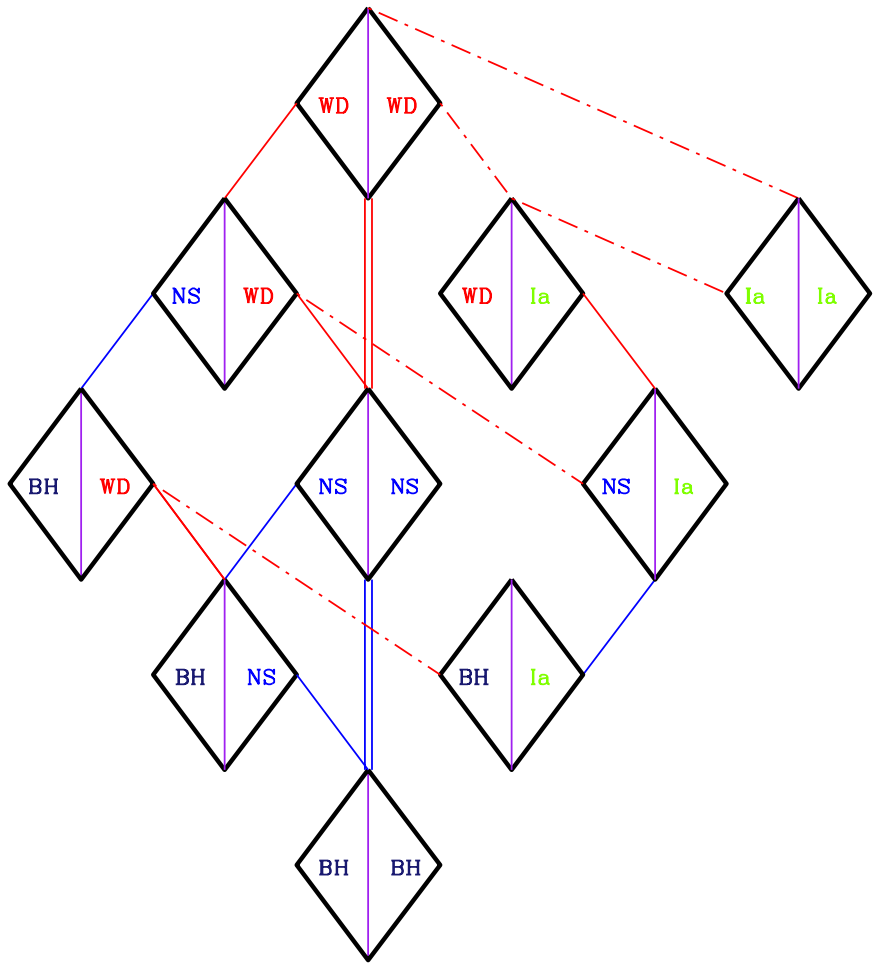}
    \caption{
Each diamond represents a close-orbit binary, and is divided into two parts, 
each representing one of the binary's components.  The links among these 
diamonds represent evolutionary pathways in which one of the binary's
components gains enough mass to make a transition that changes its nature.
Because the addition of only a relatively small amount of mass
can spark a transition from a WD to a NS or from a NS to a BH, all of these
transitions are possible.}   
    \label{fig:example_figure}
\end{figure*}

\begin{figure*}
        \includegraphics[width=2\columnwidth]{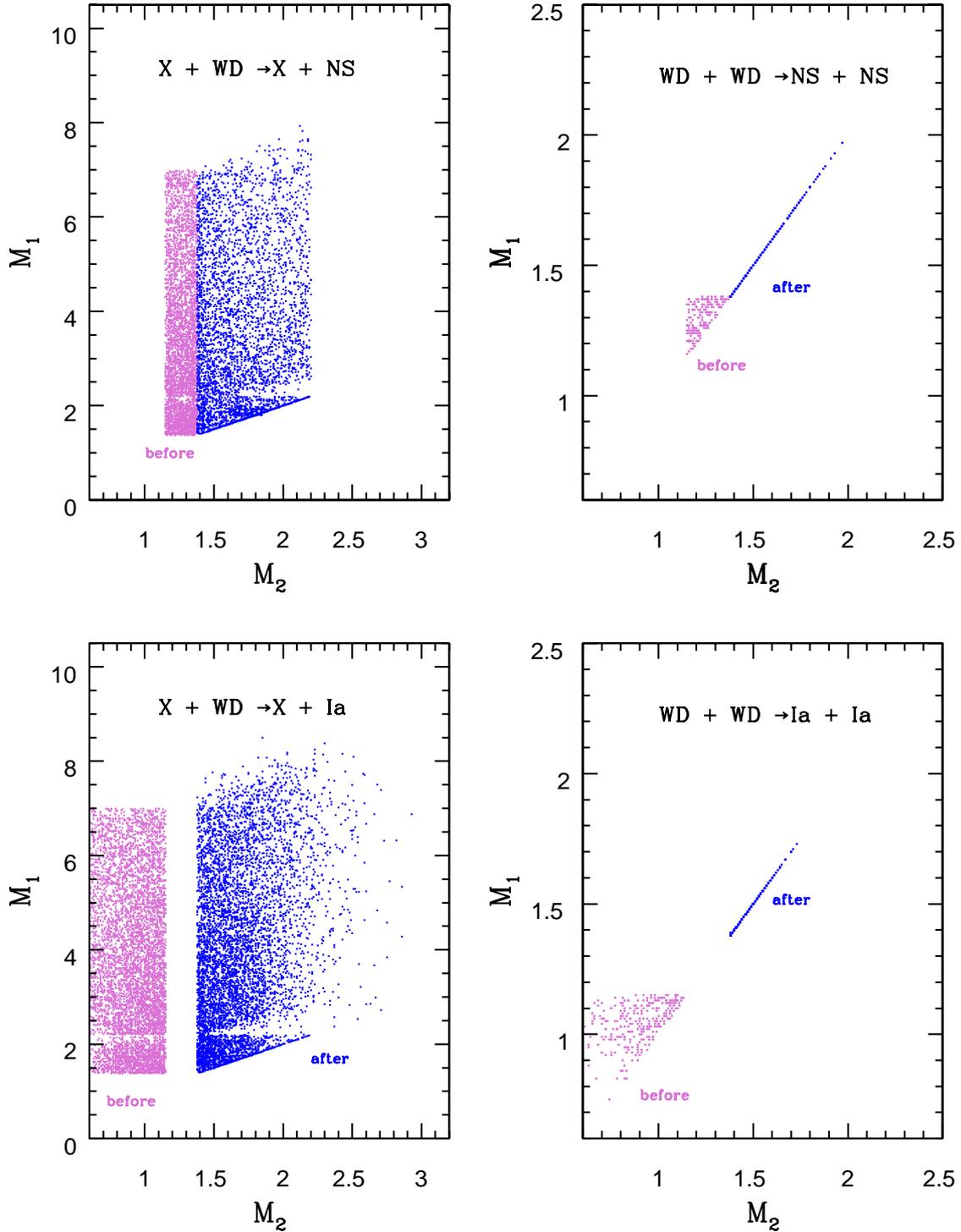}
    \caption{{\bf Initial masses (in orchid)  and final masses (in blue) 
of binaries} in which a WD undergoes
 an AIC (top panels), an SN~Ia (bottom panels). Each point 
represents a pair of masses, with the initial masses on the left (orchid points)
and the final masses on the right (blue points). The WD mass,  $M_2$, is
plotted along the horizontal axis, and its companion's mass is plotted
along the vertical axis. 
} 
    \label{fig:example_figure}
\end{figure*}
\begin{figure}
         \includegraphics[width=\columnwidth]{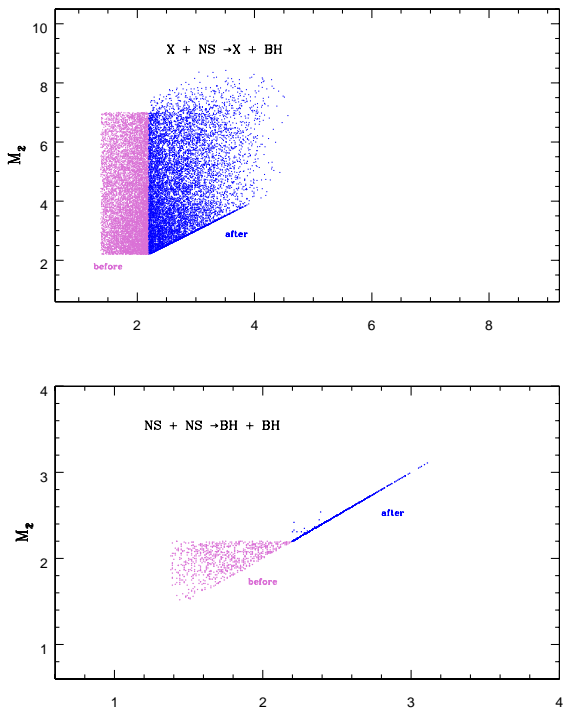}
    \caption{{\bf Initial masses (in orchid)  and final masses (in blue) 
 of binaries} in which NSs undergo
 AICs to become BHs.  In the top panel, the NS starts with a BH companion.
In the bottom panel, the NS starts with an NS  companion and both NSs collapse.
 Each point 
represents a pair of masses, with the initial masses on the left (orchid points)
and the final masses on the right (blue points). An NS mass,  $M_2$, is
plotted along the horizontal axis, and its companion's mass is plotted
along the vertical axis. 
} 
    \label{fig:example_figure}
\end{figure}

\begin{figure}
        \includegraphics[width=\columnwidth]{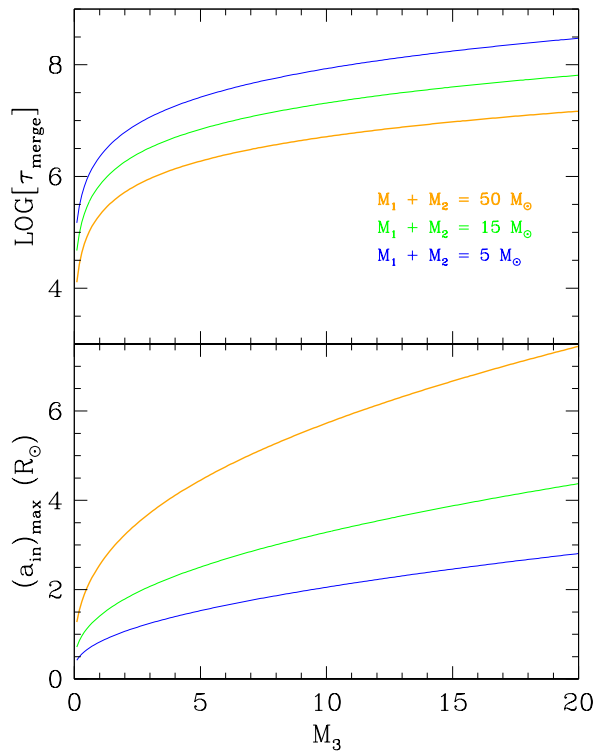}
    \caption{
{\bf Top panel: The logarithm to the base ten of the time to merger, versus the 
donor mass} of a main-sequence Roche-lobe filling donor.
Each curve corresponds to a given value of the inner 
binary's total mass, as shown in the legend. 
The value of the orbital separation of the inner binary is the maximum consistent
with the orbital stability of the hierarchical triple, ${a_{in}}_{max}$.  
{\bf Bottom panel: ${a_{in}}_{max}$
(in solar radii) is plotted versus the donor's mass.} 
}
    \label{fig:example_figure}
\end{figure}
\section{Transformations of the Natures of the Accretor(s)} 

Figure~8 illustrates the full set of transformations made possible through
the accretion of mass by one or both compact objects comprising a binary. 
Each diamond in the
figure represents a compact-object binary. The binary's components are
labeled: {\textcolor {red} {WD}}; {\textcolor {blue} {NS}}; 
{\textcolor {Violet}{BH}}.  
The least massive combination, {\color{red}{WD-WD}} 
appears at the top of the
structure, and the most massive,  
{\color{Violet}{BH-BH}},
appears at the bottom. This structure provides a simple visualization of
the possible transformative 
effects of mass infall.    

Imagine mass accreting onto the {\color{red}{WD-WD}} pair.
It may influence the orbit, shrinking the size of the diamond (which
represents the orbit) or in
some cases, expanding it. If mass stops falling before either WD reaches the
critical mass, then no transitions are made\footnote{We note, however, 
that the eventual merger of the WDs would produce an SN~Ia via the 
double-degenerate channel.}. 
If, however, one of the WDs
is an O-Ne WD that achieves the critical mass, that WD becomes a NS.
A red line connects the {\color{red}{WD-WD}} diamond to the diamond below
and to the left of it, which represents an {\color {blue}{NS}}-{\color{red}{WD}}
pair. 

Note that there are also two red lines connecting the 
{\color {red}{WD-WD}} binary to a double-NS binary. Each line represents
the AIC of a WD; two such connections side-by-side
correspond to two transitions that occur during a time interval short
compared with the time scale of mass transfer.     
While we don't expect two AICs to occur at exactly the same moment,
the near equality of the mass-gaining WDs in our accretion 
scenario suggests that two AICs
could occur very close in time to each other. 
Thus, the after effects of the first event may still be
detectable at the time of the second event. 

Every red line connecting a diamond on an upper level to 
a diamond on the level below represents
the AIC of a WD. Such transitions are expected if O-Ne WDs are in
close binaries accreting from a hierarchical companion. This is because the
amount of mass that is needed to effect the transition is at most 
$\sim 0.25\, M_\odot.$  
Thus, if O-Ne WDs are in close binaries, and also have donor stars in wide orbits,
the only condition that must be satisfied in order to produce an
AIC is that, for an interval of a few times $10^5$ years,
 mass from the donor would have to be incident on the WD
at a high enough rate ($\sim 10^{-6} M_\odot$~yr$^{-1}$) for nuclear burning to
occur, or else at the slightly lower rates that would produce recurrent novae.
This is achievable for giant donors at orbital
 separations ${\cal O}(AU)$.

Mass changes that occur when a compact object makes a transition are
illustrated in Figures~9 and 10. 
The points in Figure~9 and in Figure~10 were generated by the
full set of binaries evolved to produce Table~1. In orchid (blue) are the masses
of the WD [$M_2$] and its companion [$M_1$] prior to (after) mass transfer.

Figure~9 considers transitions of WDs. 
The top left-hand panel of Figure~9 shows cases in which a
WD collapses into a NS when its initial companion was either a NS or a BH.
In the systems shown, the companion did not transition to a compact object of 
another type. One can see that, in general, both compact objects gain mass.
The small gaps that 
appear around $M_1=2.$ correspond to systems in which the WD's
NS companion transitioned to a BH.
On the right are binaries which started with two WDs and ended with
two NSs. Note that the masses of the WDs track each
other, staying almost equal. This feature is related to our
mass-transfer scenario, in which mass flows first to the least massive component.

The two bottom panels show transitions in which one or both WDs undergo
SNe~Ia,  
Each of these panels is analogous to the panel just
above it.
The gaps in the ``before'' and ``after'' masses of $M_2$
reflect the fact that the upper mass limit for C-O WDs is
below the Chandrasekhar mass.  Otherwise the characteristics are
very similar to those illustrated above in the WD to NS transitions.

Note that we did not terminate the evolutions at the time a WD
achieves 
$M_{Ch}.$ In principle, a Type Ia supernova could occur at this point. In
practice, there will be a simmering phase prior to explosion that could last
$\sim 1000$ years \citep{2011ApJ...738L...5P}. An even longer delay is possible if 
the WD has spun up and needs to spin down to explode \citep{rd.2011}.  
Because we did not assume that a WD explodes immediately upon
reaching the Chandrasekhar mass, the masses of the WDs in the final state
exhibit a range of values, extending to above $2\, M_\odot.$ 
If the WDs are spun up by accretion, super-Chandrasekhar
explosions are expected \citep{rd.2011}. In the right-hand panel, cases in which both WDs
explode are shown. Such a scenario is possible if accretion can continue 
even after one WD has 
achieved $M_{Ch}$ (i.e., if that WD does not explode on a short time scale), 
 or else if the orbital parameters are such 
that, even after a WD undergoes supernova, its companion WD can stay bound to the donor and can continue to accrete.

In Figure~10, the top panel shows mass changes during
transitions in which a BH-NS pair becomes a BH-BH pair when the 
NS collapses. The bottom panel shows systems in which 
two NSs collapse to become two BHs. Again the near-equality
of the final BH-BH masses is exhibited. Note that 
there are clear deviations from  
exact equality. This is because the mode of mass transfer in our simulations allows mass that could not be retained by the lower-mass accretor to flow to its companion, which may then retain some of 
incident mass. 
 
{\bf Observable Signatures:} 
The question arises: how can we know if some of the compact objects we detect
are the result of transformations from less massive compact objects? Of particular relevance: how can we know that the mass that led to collapse was donated by a third star in a hierarchical triple?

The first answer is to study the distribution of masses of the compact objects
we detect through the gravitational radiation they emit during merger. Analysis
of the gravitational wave signatures allows us to measure the masses of the compact
objects that merged. While may mergers are likely to occur in systems that
did not undergo accretion from a third star, those that have followed the
evolutionary pathways we outline here could help to shape the form of the mass distribution.
For example, 
the accretion-induced collapses of NSs produced BHs that, at least at the
time of collapse, have masses lower than those of the BHs we have been studying
within X-ray binaries. Furthermore, 
if mass is channeled toward the least massive component, then the
masses of the compact objects that merge should tend to be similar. 
  Thus, our model is likely to produce pairs of merging BHs,
with the two members of the pair having similar masses, and those masses may 
each be less than $\sim 4-5\, M_\odot,$ with some hovering right near the
maximum NS mass. In addition, we may NS-BH  mergers in which the masses of
the NS and BH are nearly the same. 

WD-NS mergers will, with present-generation instruments, be detected primarily through electromagnetic signals. If, however, the mass of the WD can be established, 
our model predicts that some of the WDs 
should have masses very close to the critical mass.

AICs of WDs to NSs and NSs to BHs should also be electromagnetically luminous.
In our model. roughly half of them should take place within an accreting 
hierarchical triple. Furthermore, if the electromagnetic signature

In addition, pulsar searches are discovering many more systems in which a NS is in 
orbit with another compact companion  

\begin{figure*}
%        \includegraphics[width=\columnwidth]{/home/rd/mnras/picture.pdf}
%        \includegraphics[width=11 in]{/home/rd/mnras/picture.pdf}
%\hspace{-1.6in} 
%\vspace{-2in} 
%        \includegraphics[scale=1]{/home/rd/mnras/flow_chart_p.pdf}
         \includegraphics[scale=0.5]{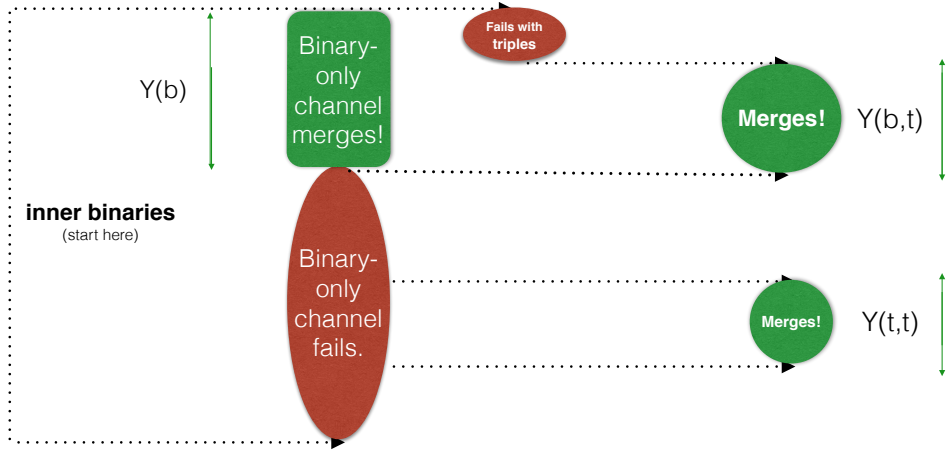}
 \vspace{-.5in} 
    \caption{
Flow chart in which all inner binaries start on the left-hand side and
evolve toward the right. 
 The green rectangle 
of length $Y(b)$ corresponds to binaries that would merge within $\tau_H$ in binary-only scenarios.
In binary-only scenarios, all other systems end at the red oval. 
In the model in which there is a 
mass-donating star in an outer orbit, all evolutions continue to the right.
Note that some systems that would have merged in the binary-only scenarios
may now fail to merge within $\tau_H$.
The flattened 
red oval represents such failures. The upper green oval on the right, 
of length $Y(b,t)$, corresponds to mergers
that could have happened, even in binary-only scenarios, and the lower green 
oval of length $Y(t,t)$ 
corresponds to mergers that are added through the effects 
of mass from the third star.
The sum of the lengths of these two ovals is $Y(b,t)+Y(t,t)$ and is
 expected to be larger than 
$Y(b)$.}   
    \label{fig:example_figure}
\end{figure*}

\section{Wider Range of Characteristics and Possibilities}

Mass transfer for a star in a wide orbit can influence the masses and orbits 
of compact objects in a close binary. We have illustrated this  
with examples of mass transfer from an evolved star, because this  process
can be followed in a simple way, closely connected to an analytic
formalism. We have focused on cases in which angular momentum from the outer
orbit is carried away from the triple.
Here we consider elements of our model 
that extend the results beyond those derived
in \S 3 and \S 4.   

\subsection{Mass Transfer from a Main Sequence Star}

Main sequence stars can 
serve as donors in hierarchical triples.
There are several important differences between systems with main-sequence and 
giant donors. 
Main-sequence stars have wind mass-loss rates typically much smaller
than evolved stars. If a main-sequence star is not filling its Roche lobe, 
the mass infall rate to the inner binary is likely to be low. We therefore
consider only cases in which the main-sequence donor fills its Roche lobe.  
The mass, $M_3$ of the main-sequence star determines its equilibrium radius, and
therefore the radius, $R_L$, of its Roche lobe.
The value of the orbital separation at which the donor
fills its Roche lobe, $a_{out}$, is determined by the  
combination of $R_L,$ $M_3$, and the total mass, $M_1+M_2$ of the
inner binary.
\footnote{At the time the donor first
comes to fill its Roche lobe, its radius is likely to be the same as the
equilibrium radius. As the star loses mass, however, 
it can either shrink or expand.
The difference from the equilibrium radius is not typically large, so
here we use it as a guide.}    

Once we know the value of $a_{out}$, we can employ the condition for 
orbital stability to determine
${a_{in}}_{max}$,
 the maximum possible value of $a_{in}$ 
 for which the triple is dynamically stable. 
The ratio $a_{out}/a_{in}$ provides a basic guideline to whether
the orbits are stable. If the effects of mass flow drive the ratio
to values that are too low, the hierarchical triple will no
longer be dynamically stable, and mergers or ejections may occur. 
If, on the other hand, the triple is dynamically stable, then 
a large fraction of the mass lost by the donor will 
come under the gravitational influence of the inner binary.

The bottom panel of Figure~11 plots
${a_{in}}_{max}$, the maximum value of $a_{in}$ consistent
with orbital stability, for three values of the total  mass, $M_1+M_2$, 
of the inner 
binary, and for a range of donor masses extending 
to $20\, M_\odot$
The upper panel plots values of the logarithm to the base 10 of
the time to merger when the separation  
is ${a_{in}}_{max}$. 

Figure~11 demonstrates that the times to merger in
all of these cases is short. For example if the total mass
of the inner binary  is $15\, M_\odot$, the time to merger ranges
from tens of thousands of years for an M-dwarf donor star to just under
$10^8$~years for a donor of $20\, M_\odot.$   
\footnote{
We used a mass-radius
relationship appropriate for stars below roughly $9\, M_\odot$.
The results would not be qualitatively different with a mass-dependent
formulation.} 
We can also consider cases in which the inner orbits are smaller than 
${a_{in}}_{max}$, since these will also be stable with respect to the
dynamical evolution. The  
times to merger would then be even shorter: in most cases shorter
than the main-sequence lifetime of the donor.

Main-sequence donors are therefore likely to
be present at the time of merger. Mass from the donor
can provide a luminous electromagnetic signature before, during, and after 
merger. The prior signal would be strong at X-ray wavelengths if one or
both of the compact objects in the inner binary accretes. The orbital period
of the inner binary may be measurable through variations in the X-ray flux.  
Furthermore, the signal from one accretor can be enhanced through 
gravitational lensing by its compact companion
\citet{2018MNRAS.474.2975D,DoDi:2018inprep}.
If the merger result and donor star are able to remain in
orbit, the long-term result after merger will be an X-ray binary.

\subsection{Angular Momentum: General Considerations}

If the angular momentum flow is more complex than in the simple examples
we have considered, there can be interesting consequences. 

\smallskip

{\bf Possible increases in the angular momentum of the inner orbit:} The 
angular momentum of the inner orbit can be significantly smaller
than the angular momentum associated with the outer orbit, yielding
small values of the following ratio.  
\begin{equation} 
\frac{L_{in}}{L_{out}} =  
\Bigg(\frac{M_1\, M_2}{M_3\, (M_1+M_2)}\Bigg)\, 
\Bigg(\frac{M_1+M_2+M_3}{M_1+M_2}\Bigg)^{\frac{1}{2}}    
\Bigg(\frac{a_{in}}{a_{out}}\Bigg)^\frac{1}{2}   
\end{equation} 
Thus, if even a small fraction of angular momentum from the outer orbit
is transferred to the inner orbit, the results can be dramatic.

Consider, for example, the case of perfectly aligned orbits in which mass 
from the outer star is accreted by one or both of the inner stars
without any mediation by, e.g., a disk. The binary can be ``spun up''.
If the triple remains stable, the time to merger will increase. 
If, however, $a_{in}$ increases and $a_{out}$
either decreases or increases at a rate slower than $a_{in}$, the triple
may become unstable, calling its final fate into question. The ejection of
one of the stars is a possibility, and so is a prompt merger. In the latter case,
the merger would take place in a mass-containing region, and the donor
star would also be present at the time of merger, potentially producing 
an electromagnetic signature.  
{\sl It is interesting to note that, if the outer orbit is not too big, expansion
of the inner orbit can trigger a dynamical instability, producing a 
prompt merger. 
}

\smallskip 
{\bf Three-dimensional rotation:}
The inner and outer orbits may not be in exact alignment. In such cases, 
mass incident on the inner binary from star~3 can induce
rotation in the
orbital plane of the inner binary. Were the inner binary isolated,  
the force of  
gravity acting on its two components would define 
a two dimensional plane. In our case, the inner orbit defines 
one plane, with an angular momentum vector $\vec L_{in}$
 perpendicular to that plane. 
The same is true of the outer orbit. The two angular momenta, $\vec L_{in}$
and $\vec L_{out}$  may point 
in different directions. Mass flow from the outer star influences not only 
the motion of the compact objects within the plane of the inner orbit, but
can also serve to rotate the orbital plane, setting it ``spinning'',
as the two compact masses continue to orbit each other. 
{\sl The complex dynamics has much in common with situations in which
mass is not flowing through the system, but instead the dynamics is dominated
by three-body interactions (second appendix).}

\subsection{Massive Donors} 

Massive donors introduce additional features. 
When such donors have close stellar companions, binary interactions
can strip them of their hydrogen envelopes \citet{1986ApJ...310L..35U}. 
Energy provided by the 
remaining nuclear-burning core can produce pre-supernova outbursts,
consistent with observational evidence for precursor events
[\citet{2018MNRAS.476.1853F} and references therein].
Heavy winds, and precursor events may inject 
mass into the orbit of the inner binary. While such mass injections may
or may not increase the mass of the inner binary, they are
likely to decrease the time to merger. Furthermore, any alterations
made close to the time of supernova would have been preceded by
an epoch of sustained winds more likely to
produce genuine mass increases and to decrease 
the orbital angular momentum. Thus, even prior to supernova, the
close binary may be more massive and closer to merger than it
would have been had the companion not been in orbit with it.

The supernova emits high-power matter  
which flows past the binary, interacting with it, possibly torquing it.  
Depending on the initial configuration of the triple, its evolution, and the response of the binary to the supernova, the binary may merge near the time 
of the supernova or at least while the supernova remnant is still
detectable.  We would then detect a    
gravitational wave (GW) source within a supernova remnant.
The supernova would not be associated with either of the merging
compact objects, however. It is therefore important to search for
tell-tale signatures, such as the location of the gravity-wave emitter
away from the center of the supernova remnant.  
Depending on the total change in the mass of the
triple-star system, the remnant of star~3 could become
unbound from the compact-object binary.

\subsection{Uncertain or New Physics}

{\bf Common Envelope:} Many elements  of the common envelope are 
not yet well understood \citet{1993PASP..105.1373I, 2018arXiv180303261M},
and references therein. These include the initiation of the common envelope; 
its evolution, and whether there is mass gain by the stars engulfed by it; 
the evolution of the stellar orbit; and the end state left behind.
In our calculations we have assumed that the compact objects in the inner 
binary gain no mass during the common envelope phase. Some evolutionary
calculations allow hypercritical accretion. (See \citet{2010CQGra..27q3001A} for
an overview for binary evolution of systems leading to 
compact binary coalescences.) Mass gain during such a phase would likely be minimal.
Nevertheless, it could be enough to change the nature of one or both components
of the compact binary.

In our conservative approach, the common envelope doesn't produce prompt
mergers and no mass is gained during the common envelope phase. 
It is likely, however, that in some real systems the consequences 
of the formation
of a common envelope are more dramatic than those allowed in our calculations.
This will have the effect of driving more close binaries to an earlier 
time of merger. 
Also, in combination with mass gained by the components of the close
binary prior to the formation of the  
common envelope, any mass gain during the common envelope could change the 
natures of some accretors.

{\bf Circumbinary Disks:}
Circumbinary disks have been studied in the context of 
supermassive black hole binaries. The least massive component of the inner 
binary passes closest
to the accretion disk, and simulations show that a {\sl minidisk}
can therefore form around it, allowing this least-massive compact object to
gain mass [\citet{2018arXiv180102266T}].
Should its mass
become equal to that of its companion, then both stars may gain mass,
with one of them accreting, and then the other. In this scenario, the
two masses stay nearly equal to each other, so that    
mass from star~3 tends to equalize the masses of the inner binary.

There are other possibilities, both for supermassive and stellar mass BHs
[\citet{2018arXiv180106179M, 2018arXiv180102624K}].
These possibilities include the 
accelerated growth of the more massive component. It is also
 worth noting that
the compact objects in the inner binary are close enough to each other
 that if
incoming matter forms a structure, such as a corona, some of the mass forming that
structure could be transferred to the other star.
This reasoning produced the flow of mass we incorporated into the
calculations of \S 3, which gave the least massive component of the
inner binary a chance to accrete incoming mass; mass which could not
be retained
by the least massive component was then passed on to the more massive 
component.

Studies of circumbinary disks show that they can
play active roles in, for example, extracting angular momentum 
from the binary. Because the possibilities are not yet well understood, 
we  have not explicitly incorporated disks into our
calculations.
Generally, a disk promotes dissipative effects. These 
tend to  
erase the effects of the detailed mass flow history.
Our approach of considering only the local flow of
matter in the vicinity of each accretor is therefore likely to be valid in
many cases.

{\bf Gravitational Lensing Within the Compact Binary:}
Consider inner binaries whose components are NSs and/or BHs. 
When mass is incident on one of these compact objects,
it emits electromagnetic  
radiation, primarily at X-ray wavelengths. 

Roughly $10\%$ of the inner binaries have orientations 
favorable for the detection of gravitational lensing. That is, the
projected distance between the luminous accretor and its companion
(during a time interval in which the companion passes in front of the accretor)  
is small enough that radiation from the accretor is significantly lensed.
(See DoDi:2018inprep, 2018MNRAS.474.2975D.)  
When this happens, the X-ray count rate is temporarily increased.
This would occur once, or (if both compact objects are accreting)
twice per orbital period.  

Furthermore, the resultant X-ray flux, produced by lensing from a baseline that
can be $10^{38}-10^{40}$~erg~s$^{-1}$, can be high. If lensing occurs,
the number of photons we receive may be large enough to permit
detection in external galaxies. Even binaries that will require more than
$\tau_H$ to merge may be detectable through such lensing.

In addition to this unique effect, which can allow the masses of the compact
objects to be determined, there are other reasons that the periodicity of the
inner orbit may be imprinted on the X-ray signal. A definitive identification of lensing would, however, nail the case for the presence of an accreting inner binary composed of NSs and/or BHs.  
The reason we have explicitly discussed NSs and BHs in this subsection, is that
the relatively larger size of WDs means that finite-lens-size effects
decrease the probability of binary-self-lensing and also diminish the
magnitude of the any effects that do occur.

\section{Summary and Implications}   

\subsection{Summary}
 
We have considered triple-star systems consisting of two 
compact objects in a close
orbit and an unevolved star in a wider orbit. 
We have shown that, for a large range of plausible initial conditions,  
mass from the outer star
can influence the evolution of the inner binary.  This is possible simply
through the agency of winds if a small fraction of the donor's mass
comes close enough to the inner binary to interact with it. Roche-lobe filling
can also play a role. If the subsequent mass transfer is stable,
 it contributes to
 the mass of the components of the
inner binary as well as possibly altering the orbital angular momentum. 
If it is unstable, then a common envelope is likely to drain angular 
momentum from the inner orbit.
The primary effects are the following.  

\begin{itemize} 

\item {\bf Increases in the mass of one or both components of the 
inner binary.}  
It is of interest that even donor stars with masses similar to or a few times
larger than that of the Sun can provide significant mass
to the components of an inner compact binary. 

\item {\bf Mass increase can transform one type of compact object into 
another,} 
with O-Mg-Ne WDs becoming NSs, and NSs becoming BHs. Such
transformations  substantially increase the pool of NS-NS,
NS-BH, and BH-BH binaries that merge within a Hubble time. 
Only a modest amount of
added mass is required to effect these transformations.    

\item {\bf Changes in the time to merger.}  These can occur
if the inner binary's components gain mass, even if its
orbital angular momentum is unchanged.
Decreases in time to merger can be more dramatic when mass from the donor
drains angular momentum from the inner orbit.

\item {\bf A new model of Type Ia supernovae.} 
When WDs that would not have merged in a Hubble time can do so because of
mass provided by a companion in a hierarchical orbit, the rates of SNe~Ia
generated through the double-degenerate channel can increase. In addition, 
mass gain by C-O WDs during mass transfer from the third star
can produce SNe~Ia through an analog of the single-degenerate model.
The first-formed WD would already have accreted matter during the
evolution of its original stellar companion that eventually produced the second WD.
Mass from the third star provides another chance for it (and a first chance for
its WD companion) to increase its mass to the Chandrasekhar mass. 
Thus, the rate of SNe~Ia through the accretion channel can also be increased
because of mass provided by the outer star.  

\smallskip

\end{itemize}

\noindent Our calculations have been conservative. It therefore seems likely that
gains in mass and losses of orbital angular momentum much
  larger than those we have considered are possible. 
For example, {\bf (a)}~the third star could be more massive 
than the stars we considered,
{\bf (b)}~the ejection of 
the common envelope almost certainly drains more angular momentum than
we have assumed, {\bf (c)}~there may be more than one outer-orbit star that
can come to donate mass.

\subsection{Implications}

Our model has several points of connection to observations. Key elements of it can 
therefore be tested.  

\subsubsection{Gravitational Mergers}

Triples in which the outer star sends mass toward the inner binary
can increase the rate of gravitational mergers. 
\footnote{In an appendix we briefly 
discuss the role of three-body dynamics, even when mass is not transferred from the outer star.} 
Consider, for example,
BH-BH mergers. Binary interactions provide a baseline prediction. 
Binary-only calculations
apply to true binaries and to hierarchical triples in which the third star 
is too distant to influence the fate of the inner binary.
We have shown that interactions with mass provided by the outer star 
can change the time-to-merger from values larger than $\tau_H$ to 
values smaller than $\tau_H$.  
This adds to the total reservoir of binaries 
that can merge within a Hubble time. Note that the number of binaries 
added to 
this reservoir is almost certainly much 
larger than the numbers that leave it by being pushed to longer merger times by the effects of mass infall from an outer star. 
In addition, the conversion of NSs to BHs, 
and even of WDs to NSs and then to BHs
provides a brand new reservoir of binaries that will experience BH-BH mergers.
If triple systems are not rare,
such transformations have a good chance  
of contributing significantly because binaries containing the less massive
stars that produce NSs and WDs are more common than the binary systems
producing only BHs. There are caveats, including the effects of supernovae.
Nevertheless it is worth noting that binaries undergoing transformations 
should be considered as potentially important sources of BH-BH mergers. A parallel
set of arguments applies to mergers involving NSs. 

In addition to influencing the merger rates, mass from a third star
can change the distribution of merger properties.  
In particular, the merging components are likely to have nearly equal mass,
at least if the principles we have used to trace the path of incoming mass are correct.
The values of the masses depend on how much mass is available from the third
star and on the efficiency of accretion.  
If the total mass added to the inner binary is small, then 
we expect there to be merging NSs with masses near the Chandrasekhar mass, 
and low-mass merging BHs, which masses less than about $5\, M_\odot.$
If the outer star is more massive, and/or of there is a sequence of 
outer stars,   
mergers of nearly-equal high-mass BHs are expected.  

\subsubsection{X-ray emission} 
Continuing accretion 
is also a signature expected for a subset of our post-merger
systems. 
  If the compact object formed 
through the merger continues to accrete material provided by 
the donor star, it will emit X-rays.  Whether we would be able
to detect the X-rays depends of the accretion luminosity and the 
distance from us. Consider, for example, an X-ray source with
X-ray luminosity $L_X=10^{40}$~erg~s$^{-1}$, and a power-law spectrum with
$\Gamma=1.7$ and intervening gas with $N_H = 1 \times 10^{21}$~cm$^{-1}$. If
this X-ray source (XRS) were $50$~Mpc away,  
{\sl Chandra's} ACIS-S detector would record roughly $2$ counts per ks,
so that the source would be detectable. The proposed {\sl Lynx} mission
could record $\sim 50$ times as many counts, potentially
allowing short-time-scale
variations to be traced. Thus, 
post-mergers at intermediate distances may be detected as X-ray binaries.

{\bf Hierarchical X-Ray Triples} 
Finally, many of the types of systems we consider should
be detectable prior to merger as X-ray hierarchical triples. 
In our model of mass impinging on an inner binary,
the compact accretors may be detected as powerful X-ray emitters. 
X-ray emission therefore provides a way 
to identify systems within which a compact-object binary is accreting.
The signature for which in order to identify hierarchical triples
 is periodic or quasiperiodic modulation
of the X-rays, where the repetition times are harmonics of the
inner orbital period. There are many possible 
reasons for X-ray from binaries to exhibit a range of
periodic and/or quasiperiodic signatures: for example, complex accretion flows
and warped disks can introduce periodicity.
It is therefore 
important to model the signatures in order to test
whether they are consistent with accretion onto an inner binary, or whether
another explanation is equally good or better.   

Typical galaxies house dozens of bright X-ray binaries. 
If accretion from a wide-orbit star is common, then a fraction
of X-ray sources that we have assumed to be binaries may actually be
hierarchical triples. Even if there are only a handful of X-ray triples 
among the hundreds of bright X-ray sources in nearby galaxies,
it may be possible to identify them in archived data.
We note that only a fraction of the triples may include 
close binaries    
that will merge in a Hubble time. Thus, the numbers of hierarchical triple
accretors could be relatively large compared with the lifetime-scaled merger rate.
The discovery of such systems would be a powerful indication that our model
of mass transfer to compact inner binaries may be important.

\subsection{Conclusions}

Close binary systems form one of the most important classes of astrophysical
objects. Their importance is highlighted by the fact that aLIGO has
begun to discover the gravitational-wave signatures of their mergers.
This makes them the first observed multimessenger emitters.
We have introduced triple-star pathways that have the potential to contribute
substantially to the rates of gravitational mergers. This  reduces the
pressure on other channels to produce all events.  The contributions of
triples however, have still to be quantified relative to the contributions of
isolated binaries. This will be an important next step. The ubiquity
of triples involving massive stars tells us that it is an important
channel to explore.
It remains to quantify the relative contributions of hierarchical triples to several important processes: Type Ia supernovae,  merging NS-NSs, NS-BHs, and BH-BHs.
Do triples contribute only a small fraction? Or do they make significant and measurable differences to the event rates?

These questions can be answered in part through continuing observations of the events themselves. For example, are merging low-mass BHs more common than expected based on binary interactions alone? Other observations, for example improved measurements of the properties of primordial triples, will also be important.

Theoretical work is also needed to determine the role that 3-body dynamical interactions have in helping to determine the initial characteristics of inner binaries: what are the initial conditions needed as input to calculations designed to realistically model mass transfer from a third star? Other questions have to do with developing more detailed evolutions of mass transfer from a third star. For example, how are winds focused? What is the role of irradiation? What is the geometry of accretion?

We have shown that  the conditions needed for components of an inner binary 
to interact with mass from a star in a hierarchical orbit should be common
in that they extend 
over a wide range of orbital separations, donor masses, and 
characteristics of the inner binary. Accretion from an outer star is, 
therefore,  a process that is certain to occur. 
Fortunately there are many links between our model and a
range of observational signatures that may be detected pre-merger, during
merger, or even in the epoch after merger.
In addition there are connections to Type Ia
supernovae.

\section{Appendix: Roche-lobe approximation}

Mass from star 3 can be channeled directly to the inner binary in a process
analogous to what happens when a donor star in an isolated binary system 
fills its Roche lobe. 

The definition of the Roche lobe of
a star in a binary is based on the existence of an equipotential 
surface that takes into account both the
gravitational attraction of both the donor and its companion, and 
the rotation of the donor (whose
spin period is equal to the orbital period). This equipotential 
surface is smooth and constant in a frame that
rotates with the outer binary. Our hierarchical triple, however, 
introduces a crucial new feature:
a time-dependent gravitational potential, 
$\Phi(t)$, whose instantaneous value
at any point in space depends of the positions of the individual 
components of the inner binary.   

Fortunately, the fact that the outer orbit is significantly wider 
than the inner orbit means that
a good approximation to the gravitational potential of its donor 
can be written as a 
sum of (a)~a time-independent monopole term,
corresponding to the concentrating the total mass of the inner 
binary at its center of mass; 
and (b)~a time dependent dipole term associated with the orbital 
motion of the inner binary.  
The dominant term is the monopole term, which is the same as it 
would be for a single star with 
mass $M_T = M_1+M_2$. 
In addition, the donor star in the hierarchical triple would have 
a rotational period 
roughly equal to its orbital period, although this too is an
approximation, since the motion of the individual stars in the close binary
can introduce time-dependent tidal interactions.       
Thus, the Roche-lobe formalism can be applied to the case of mass 
transfer from a wide-orbit star
onto a close-orbit binary, although  
there may be observable signatures associated with the 
short-orbital-period binary.    

\section{Appendix: Dynamical three-body interactions}

The presence of the third body may have played a role 
in the earlier evolution of the hierarchical triple. For example, the Lidov-Kozai 
mechanism creates an interplay between the eccentricity of the inner orbit
and its orientation relative to the outer orbit 
\citet{1962AJ.....67..591K, 1962P&SS....9..719L, 2016ARA&A..54..441N}. 
A change in eccentricity 
can promote mass transfer, influencing the evolution of the inner binary. 
Thus, the initial conditions for mass transfer from the outer star 
may have been influenced by prior 3-body interactions.
These dynamical interactions could have been active before the components of the 
inner binary interacted; during the binary interactions that 
produced the close-orbit compact-object 
binary\footnote{Dynamical interactions with the third body 
are not likely to play a dominant role during the intervals of most 
active mass transfer. 
They could, however, serve to enhance mass transfer. In addition, there are intervals,
e.g., after the more massive of the two stars star has evolved and before its
close companion evolves, when dynamical interactions can be relatively important.}; 
and/or after the close binary
was formed. It will be important to consider the full range of prior interactions
in order to develop the profile of realistic configurations for the 
start of mass transfer from the third star. 

Furthermore, if the inner binary has not yet merged by the time mass transfer has 
ceased, 
3-body dynamical interactions continue to influence the characteristics of the inner orbit.
These interactions have the potential to influence the time at 
which the eventual merger will occur, and should be carefully considered.

\section*{Acknowledgements}
%The Acknowledgements section is not numbered. Here you can thank helpful
%colleagues, acknowledge funding agencies, telescopes and facilities used etc.
I would like to thank 
Daniel D'Orazio for discussions, and he, 
Morgan McCleod, and Amber Medina for 
their careful reading and comments on the manuscript.  
%%%%%%%%%%%%%%%%%%%%%%%%%%%%%%%%%%%%%%%%%%%%%%%%%%



\bibliographystyle{mnras}
\bibliography{tmt_39,alison_update} % if your bibtex file is called example.bib

%\begin{thebibliography}{99}
%\end{thebibliography}

%%%%%%%%%%%%%%%%%%%%%%%%%%%%%%%%%%%%%%%%%%%%%%%%%%

\end{document}